\documentclass[aps,prd,twocolumn,groupedaddress,nofootinbib,superscriptaddress]{revtex4-1}
\usepackage{amsmath, amssymb}
\usepackage{mathtools}
\usepackage[pdftex]{graphicx}
\usepackage[colorlinks]{hyperref}
\usepackage{color}
\usepackage{setspace, enumitem}
\usepackage[parfill]{parskip}
\usepackage[normalem]{ulem}
\usepackage{enumitem}
\usepackage{physics}
\usepackage[dvipsnames]{xcolor}
\usepackage{simpler-wick}
\usepackage{float}
\usepackage{amsmath}
\usepackage{hyperref}
\usepackage[capitalise]{cleveref}
\usepackage{verbatim}
\graphicspath{ {plots/} }

\newcommand{\fstop}{\, .}

\newcommand{\gagg}{g_{a\gamma\gamma}}
\newcommand{\gann}{g_{aNN}}
\newcommand{\gae}{g_{ae}}

\newcommand{\datir}{\mathcal{D}_{\text{IR}}}
\newcommand{\datuv}{\mathcal{D}_{\text{UV}}}

\newcommand{\cag}{c_{a\gamma}}
\newcommand{\cau}{c_{au_i}}
\newcommand{\cad}{c_{ad_i}}
\newcommand{\calep}{c_{aL_i}}

\newcommand{\emano}{\mathcal{E}}
\newcommand{\cano}{\mathcal{N}}

\begin{document}
%=============================================================================
\title{Probing the UV with IR Axion Dark Matter Experiments}

\author{Jacob M. Leedom}
\email{leedom@fzu.cz} % 0000-0003-4911-2188
\affiliation{CEICO, Institute of Physics of the Czech Academy of Sciences\\
Na Slovance 2, 182 00 Prague 8, Czech Republic}

\author{Serah Moldovsky}
\email{serah\_moldovsky@berkeley.edu } 
\affiliation{Theory Group, Lawrence Berkeley National Laboratory, Berkeley, CA 94720, USA}
\affiliation{Berkeley Center for Theoretical Physics, University of California, Berkeley, CA 94720, USA}

\author{Hitoshi Murayama}
\email{hitoshi@berkeley.edu } 
\affiliation{Theory Group, Lawrence Berkeley National Laboratory, Berkeley, CA 94720, USA}
\affiliation{Berkeley Center for Theoretical Physics, University of California, Berkeley, CA 94720, USA}
\affiliation{Kavli Institute for the Physics and Mathematics of the Universe (WPI),
University of Tokyo, Kashiwa 277-8583, Japan}

\date{\today}

%=============================================================================

\begin{abstract}
There is an ongoing and future experimental program searching for axions as both extensions of the Standard Model and as candidates for the observed dark matter. 
In this paper, we examine the feasibility of being able to pin down precise features of axions within the context of this program. We define a set of infrared data that can be probed by experiments and explain how it encodes the ultraviolet data defining particular models of the QCD axion. We then critically examine the extent to which these data can be determined. If the parameters are favorable, we will be able to determine the axion couplings to photons, electrons, nuclei, as well as the dark matter halo density. However, further experimental proposals are needed to fully flesh out the ultraviolet model of the axion.
\end{abstract}

\maketitle

%----------------------------------------
\section{Introduction}
\label{sec:intro}
%----------------------------------------
Of the potential extensions to the Standard Model of particle physics (SM), the QCD axion~\cite{Weinberg:1977ma,Wilczek:1977pj} remains one of the most well-motivated candidates. In addition to solving the Strong CP problem via the Peccei-Quinn mechanism~\cite{Peccei:1977ur,Peccei:1977hh}, the QCD axion has natural production mechanisms that can explain the current dark matter density~\cite{Preskill:1982cy,Abbott:1982af,Dine:1982ah}.

Proof of the existence of a QCD axion, as dark matter or otherwise, is sought  in experiments through axion-SM interactions. As a consequence of their (approximate) shift symmetries, axions generically couple to fermions and gauge bosons via derivative interactions and Chern-Simons terms, respectively. The ambitious ongoing and upcoming experimental program of axion dark matter searches can be broadly categorized into two groups -- haloscopes, instruments that probe the dark matter halo, and helioscopes, instruments that search for axions emitted by the Sun. These experiments can be further stratified by specifying the specific coupling that is utilized for a signal -- for example, light-shining-through-wall experiments probe the axion-photon coupling.  
 
The detection of an axion signal in one of these experiments would yield not only a monumental discovery in the form of a new fundamental particle, but also insight into properties of the axion and the underlying structure that lies beyond the Standard Model. 
Ultraviolet (UV) completions of QCD axion models typically involve either a spontaneously broken global U(1) symmetry or extra dimensions. Both of these options can have a multitude of variations, giving rise to a veritable landscape of QCD axion models -- see~\cite{DiLuzio:2020wdo} for a comprehensive review.

Given this landscape of models, a natural and critical question is: to what extent will the axion experimental program be able to unravel the true UV model of the axion effective field theory (EFT), and how quickly will we be able to do so?

In this work, we address these questions by analyzing the present and future QCD axion experimental program. We approach this by defining a set of minimal IR data and determining to what extent this encodes data of the UV model. We also discuss necessary steps and needed future experiments to completely pin down the true QCD axion model of the Universe, if it exists. We primarily focus on the photon coupling of the axion because of near-future progress in ongoing experiments. However, axion models are defined by more data than just the photon coupling, and so we will also examine the implications of future nucleon and electron axion experiments. 

In~\cref{sec:axions}, we outline the data required to define an infrared model of QCD axion dark matter as well as reviewing various ultraviolet models that lead to different IR predictions. In~\cref{sec:modeldisc}, we outline the data obtained from various types of axion experiments as well as several potential discovery scenarios that would allow for extraction of different elements of the IR data. In~\cref{sec:axpro} we analyze this scenarios through the lens of concrete proposals in the axion experimental program and determine realistic expectations for how experiments will fare in determining the features of QCD axion models. Concluding in~\cref{sec:conclusion}, we reflect on future prospects for axion experiments. 

%%%%%%%%%%%%%%%%%%%%%%%%%%%%%%%%%%%%%%%%%%%%%%%%%%%%%%%%%%%%%%%%%%%%%%%%%%%%%%%%%%%%%%%%%%%%%%%%%%%%%%%%%%%%%%%%%%%%%%%%%%%%%%%

%%%%%%%%%%%%%%%%%%%%%%%%%%%%%%%%%%%%%%%%%%%%%%%%%%%%%%%%%%%%%%%%%%%%%%%%%%%%%%%%%%%%%%%%%%%%%%%%%%%%%%%%%%%%%%%%%%%%%%%%%%%%%%%
\section{The Data Defining QCD Axion Dark Matter}
\label{sec:axions}

In this section, we consider the IR and UV data describing dark matter models of the QCD axion. We do this by defining two distinct sets, $\datir$ and $\datuv$. $\datir$ encapsulates the IR data, i.e. model parameters that can be probed directly by experiments, such as couplings. On the other hand $\datuv$ is the set of data that defines the UV completion of the QCD axion model. These are parameters such as anomaly coefficients or the size of cycles in a compactification manifold. We start with describing $\datir$ and then explain how its elements encode those of $\datuv$. We also review standard axion models and how they fit into this classification.

From the perspective of effective field theory, axions are pseudo-scalar particles with approximate continuous shift symmetries. From the UV perspective, axions arise as pseudo-Nambu-Goldstone bosons from spontaneous symmetry breaking or as zero modes from the dimensional reduction of $p$-form gauge fields. The shift symmetries are broken by non-perturbative effects and/or by highly suppressed higher-dimensional operators, but the approximate symmetries restrict the lowest-dimensional  operators in the EFT Lagrangian. In particular, an axion $a$ can couple to fermions and gauge fields via the operators
\begin{equation}
    \mathcal{L} \supset %\frac{C_{\Psi}}{2f_a}
    \frac{1}{2}g_{a\Psi}
    (\partial_\mu a)\bar{\Psi}\gamma^\mu \gamma_5 \Psi + \frac{1}{4} g_{agg}a G^a_{\mu\nu}\widetilde{G}^{a\mu\nu} 
\label{eq:axioncoup1}
\end{equation}
where $\Psi$ is a Dirac fermion, $G^a_{\mu\nu}$ is the field strength of a Yang-Mills gauge field, and  $\widetilde{G}^{a\mu\nu} = \frac{1}{2}\epsilon^{\mu\nu\rho\sigma}G^a_{\rho\sigma}$ is the dual field strength. Multiple copies of~\cref{eq:axioncoup1} can exist in the EFT Lagrangian if the axion couples to multiple fermions and/or gauge fields. The constants $g_{a\Psi}$ and $g_{agg}$ determine the coupling strength of the axion and are therefore elements of the set of IR data $\datir$. 

The continuous shift symmetry of the axion is preserved by the fermion coupling in~\cref{eq:axioncoup1} but in the presence of instantons, the Chern-Simons coupling breaks it to a discrete symmetry and introduces a mass $m_a$ for the axion. Other non-perturbative effects can also generate axion masses, as discussed below. However, we are primarily interested in QCD axion models, which by definition must have their mass primarily arising from a Chern-Simons coupling to QCD and the subsequent non-perturbative effects. We will describe this in more detail below, but for now we simply append $m_a$ to the list of IR data $\datir$.

The QCD Chern-Simons coupling 
required for the QCD axion mass will additionally couple to  hadrons, leptons, and photons due to loop effects, even if it does not couple to quarks at tree level via the terms in~\cref{eq:axioncoup1}. Thus the EFT of the QCD axion will have interaction terms
\begin{equation}
    \begin{split}
    \mathcal{L} &\supset 
    \frac{2g_{a\pi}}{f_{\pi}}
    \partial_\mu a\bigg(2(\partial^\mu\pi^0)\pi^+\pi^- - \pi_0(\partial^\mu\pi^+)\pi^--\pi_0\pi^+\partial^\mu\pi^-\bigg)\\
               &\hspace{1.2cm} + 
               \frac{g_{an}}{2}(\partial_\mu a)\bar{n}\gamma^\mu \gamma_5 n+
               \frac{g_{ap}}{2}
               (\partial_\mu a)\bar{p}\gamma^\mu \gamma_5 p\\
               &\hspace{1.2cm}+
               \frac{g_{ae}}{2}
               (\partial_\mu a)\bar{e}\gamma^\mu \gamma_5 e + \frac{1}{4}\gagg a F_{\mu\nu}\widetilde{F}^{\mu\nu} + \cdots
    \end{split}
\label{eq:axionSMcouplings}
\end{equation}
The $\pi$ fields correspond to the pions, while $n$, $p$ and $e$ represent the neutron, proton, and electron spinors, respectively. $F_{\mu\nu}$ is the standard  electromagnetic field strength tensor. The couplings $\{g_{a\pi}, g_{an}, g_{ap}, g_{ae}, \gagg,\cdots \}$  constitute further elements of the IR data $\datir$.

Finally, a key coupling of the QCD axion involves the operator
\begin{equation}
    \mathcal{L} \supset -\frac{i}{2}g_d a\bar{N} \sigma_{\mu\nu}\gamma_5N F^{\mu\nu}
\label{eq:axionEDM}
\end{equation}
where $N$ is a nucleon spinor 
is the spinor for a nucleon, $\sigma_{\mu\nu}$ is the usual Dirac matrix bilinear, and $F^{\mu\nu}$ is the electromagnetic field strength tensor. This term arises from the direct coupling of the axion to the gluon via the Chern-Simons interaction $aG\widetilde{G}$. While the coupling constant in~\cref{eq:axionEDM} does not introduce a new IR parameter, measuring this operator is nonetheless essential since it is the only definitive method to differentiate a QCD axion from an axion-like-particle (ALP).

The above parameters collected together form a subset of $\datir$. They specify the particle model details for a QCD axion model, but they are not sufficient to completely determine a \textit{dark matter} model of the QCD axion. At a minimum, we must supplement the above list with the dark matter density $\rho_{DM}$ and the dark matter velocity distribution $f(v)$. With these, the data defining the IR QCD axion model is
\begin{equation}
    \datir = \{m_a, \rho_{DM}, f(v), \gagg,g_d, g_{a-},\cdots\}
\label{eq:IRDATA}
\end{equation}
where $g_{a-}$ stands for the coupling constants of the axion to SM fermions, mesons, and baryons. The dots indicate that there are potential other additions to this set -- for example, one could also include axion coupling to the Higgs, but we will largely be concerned with only a subset of this data and restrict $g_{a-}\in \{g_{ae},g_{an},g_{ap} \}$. We also note that one could include other parameters detailing the dark matter production mechanism, such as the initial displacement angle for the misalignment mechanism. However, we will treat these as UV parameters since they are not directly probed by experiments, unlike the density and velocity distribution.  

We now move onto describing the set of UV data $\datuv$ and how to extract values for its elements from $\datir$. The most straightforward UV parameter associated with axion models is the axion decay constant $f_a$, which appears in the mass and couplings. To see this, we recall from above that axions can obtain masses from a variety of non-perturbative UV effects - examples include gauge instantons, gravitational instantons from a coupling $aR\widetilde{R}$, euclidean D-branes, and gaugino condensation. 
All these effects a commonly modeled via a periodic potential of the form\footnote{The precise form from the chiral Lagrangian is more complicated, but the rest of the discussion is not affected significantly.}
\begin{equation}
    V(a) = \Lambda^4\bigg(1- \cos\left(\frac{a}{f_a}\right)\bigg) 
\label{eq:axpot}
\end{equation}
The overall coefficient $\Lambda$ may depend on other fields in the UV model, but we will take it to be a constant representing the scale of the non-perturbative physics. Clearly~\cref{eq:axpot} gives rise to an axion mass $m_a = \frac{\Lambda^2}{f_a}$. This expression illustrates how two UV parameters, $\Lambda$ and $f_a$, are encoded in the IR parameter $m_a$. However, for our purposes, we will need only $f_a$. For the QCD axion, the mass arises primarily from QCD instanton effects, as described above. In this case, the axion mass is then a simple function of $f_a$~\cite{diCortona:2015ldu}:
\begin{equation}
    m_a \simeq 5.7\bigg(\frac{10^{12}\text{ GeV}}{f_a}\bigg)\;\;\mu\text{eV}\fstop
    \label{eq:mass-faRelation}
\end{equation}
Thus we will take $f_a$ as an element of $\datuv$. The other IR couplings discussed above can similarly be recast in term of UV data. we note that the Chern-Simons coupling to photons as displayed in~\cref{eq:axionSMcouplings} can be expressed as 
\begin{equation}
\gagg = \frac{\alpha_{EM}}{2\pi f_a}\bigg(\cag^0- \frac{2}{3}\frac{4m_d+m_u}{m_u+m_d}\bigg)
\label{eq:QCDphoton}
\end{equation}
Here $\alpha_{EM}$ is the electromagnetic fine structure constant. The second term in parentheses, which depends on the up- and down-quark masses, is induced from the QCD Chern-Simons term by mixing with the neutral pion. The first term consisting of $\cag^0$ represents the direct coupling of the axion to electromagnetism. In simple UV completions of the QCD axion, $\cag^0 = \mathcal{E}/\mathcal{N}$, where $\mathcal{E}$ and $\mathcal{N}$ are the electromagnetic and color anomaly coefficients, respectively. In non-trivial models, $\cag^0$ can deviate from this expression by enhancement/suppression factors. Thus in general we set
\begin{equation}
    \cag^0 = \sigma \frac{\mathcal{E}}{\mathcal{N}},
    \label{eq:directaxionphotonc}
\end{equation}
with $\sigma=1$ for simple QCD models. We therefore include $\{\sigma, \mathcal{E}, \mathcal{N}\}$ in $\datuv$. Moving onto the fermion, meson, and baryon couplings, we first note that the direct couplings to fermions can be expressed in terms of $f_a$:
\begin{equation}
    g_{a\Psi} = \frac{c_{a\Psi}}{2f_a}
\label{eq:IRcoup}
\end{equation}
This holds for the electron, proton, and neutron couplings in~\cref{eq:axionSMcouplings} as well as the pion couplings. For the electron, $c_{ae}$ represents the direct coupling of the axion to the electron and is an element of $\datuv$. For the proton and neutron, more complicated expressions hold. For example,
\begin{equation}
c_{ap} = - \bigg(\frac{m_d}{m_u+m_d}\Delta u + \frac{m_u}{m_u+m_d}\Delta d\bigg) +c_{au}^0\Delta u + c_{ad}^0\Delta d
\end{equation}
The $\Delta u$ and $\Delta d$ are numerical coefficients and $c_{au}^0$ and $c_{ad}^0$ are the tree-level coupling of the axion to up and down quarks, respectively. Thus the UV parameters of interest for the proton and neutron couplings are the axion-quark couplings. For the meson couplings, the expressions in~\cref{app:couplings}, one sees that the meson interactions are also determined by tree-level axion-quark interactions. Finally, the EDM coupling in~\cref{eq:axionEDM} is determined by the axion decay constant as $g_d\propto (m_Nf_a)^{-1}$.

Utilizing the above, we see that the UV data that gives rise to $\datir$ can be written as
\begin{equation}
\datuv \supset \{\sigma, \Lambda, \emano, \cano,  f_{PQ}, \cau^0,\cad^0,\calep^0,\theta_a\}
\label{eq:UVDATA}
\end{equation}
The index $i=1,2,3$ denotes the generation of the axion-fermion coupling. We have also replaced $\rho_{DM}$ with data defining the production mechanism of the axion dark matter, denoted collectively as $\theta_a$. Leveraging experiments to determine $\theta_a$ is an interesting and non-trivial task. However, in this work, we will largely be concerned with terrestrial experiments and so unraveling $\theta_a$ will be left for future studies. Note that the above data can also be recast, particularly in the context of string compactifications. For example, the axion decay constant can be translated for the size of certain submanifolds in the compactification manifold~\cite{Svrcek:2006yi}.
%==============================================================================================================================
\subsection*{QFT Models of the QCD Axion}
In this subsection, we review models of the QCD axion as an illustrations of concrete examples of $\datuv$. We primarily consider the Dine-Fischler-Srednicki-Zhitnitsky (DFSZ)~\cite{Zhitnitsky:1980tq,Dine:1981rt} and the Kim-Shifman-Vainshtein-Zakharov (KSVZ)~\cite{Kim:1979if,Shifman:1979if}. In both models, the axion arises from the phase of a complex scalar field $\Phi$ whose vev $\langle\Phi\rangle = f_{PQ}$ breaks a global U(1)$_{PQ}$ symmetry. However, the models differ in key aspects and yield different values for the elements of $\datuv$.

In the KSVZ model, the Standard Model particle content is supplemented by $\Phi$ and additional fermions $\{Q,\bar{Q}\}$ that are charged under QCD. The axion arises as the pseudo-Nambu-Goldstone boson associated with the phase of $\Phi$ and does not have any direct, tree-level coupling with the Standard Model. Thus in terms of the data in $\datuv$, $c^0_{a\Psi}=0$ for all SM fermions $\Psi$. The KSVZ axion will couple to SM gauge fields via Chern-Simons terms arising from the new quarks. The anomaly coefficients depend on the charges and number of new fermions. Thus for KSVZ axions, we have
\begin{equation}
    \begin{split}
        \datir^{KSVZ} &\supset \{g_{an} = -0.02(3), g_{ap} = -0.47(3)\}\\
        \datuv^{KSVZ} &\supset \{ c^0_{u^i}=0, c^0_{d^i}=0,c^0_{L^i}=0        \}
    \end{split}
\end{equation}
For KSVZ-like models, the anomaly coefficients $\emano$ and $\cano$ depend on the number and representations of new quark states.

\begin{figure}
    \centering
{}    \includegraphics[width=\columnwidth]{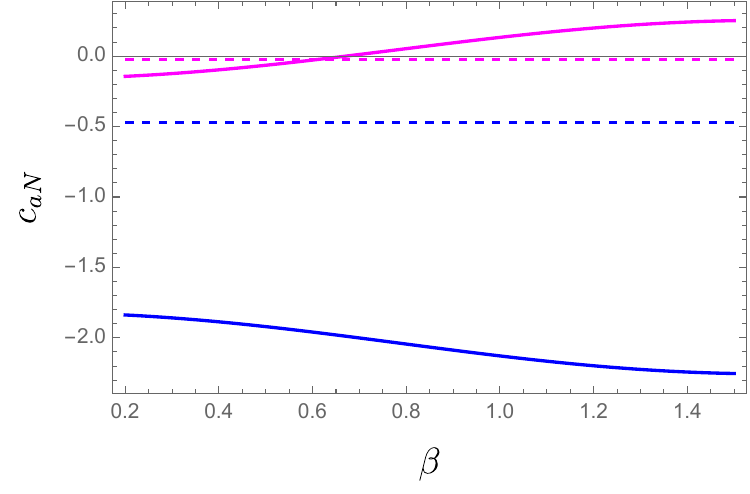}
    \caption{$c_{an}$\& $c_{ap}$ for KSVZ and DFSZ Axion Models. $c_{ap}$ is plotted in blue with the solid line the value for DFSZ and the dashed for KSVZ. $c_{an}$ is shown in magenta for DFSZ (solid) and KSVZ (dashed).}
    \label{fig:nucleoncoup}
\end{figure}

On the other hand, the DFSZ model extends the SM particle content with $\Phi$ and an additional Higgs doublet so that the Higgs content consists of $H_u$ and $H_d$. One distinguishes between two subclasses of DFSZ models depending on whether the leptons couple to $H_d$ (DFSZ-I) or $H_u$ (DFSZ-II). The QCD axion is now a linear combination of the Nambu-Goldstone bosons arising from the scalar fields and therefore couples directly to SM fields. These couplings depend on the mixing angle $\beta$, which is constrained~\cite{DiLuzio:2016sur,DiLuzio:2017chi,Bjorkeroth:2019jtx} 
\begin{equation}
    0.25<\tan\beta < 170 
\label{eq:betas}
\end{equation}
for both DFSZ variants.  The UV data of the two variants does differ:
\begin{equation}
    \begin{split}
        \datuv^{DFSZ-I} &= \bigg\{\frac{\emano}{\cano} = \frac{8}{3}, c^0_{u_i} = \frac{1}{3}\cos^2\beta, c^0_{d_i} = \frac{1}{3}\sin^2\beta, \\
        &\hspace{1cm} c^0_{L_i} = \frac{1}{3}\sin^2\beta \bigg\}\\
        \datuv^{DFSZ-II} &= \bigg\{\frac{\emano}{\cano} = \frac{2}{3},c^0_{u_i}= \frac{1}{3}\cos^2\beta, c^0_{d_i} = \frac{1}{3}\sin^2\beta, \\  &\hspace{1cm} c^0_{L_i} = -\frac{1}{3}\cos^2\beta    \bigg\}
    \end{split}
\end{equation}
We will not write out the corresponding IR data entries for these models, but they follow from the expressions above. Much of our focus will be on distinguishing between these two models, as shown in Fig. ~\ref{fig:nucleoncoup}. For $g_{an}$ there is a value of $\beta$ where the couplings of the two models overlap. But for $g_{ap}$ there is no overlap, so a measurement of this coupling could be used in distinguishing between these two types of axions.

The above two models are highly attractive due to their simplicity. However, there are a number of interesting extensions that lead to different predictions. A common model-building goal has been to boost/suppress the axion couplings for a given $f_a$. For example, naively 
the photon-axion coupling strength is $\simeq \mathcal{O}(1)/f_a$. This is due to the understanding that the Chern-Simons coupling depends on the ratio of anomaly coefficients. However, with some non-trivial model building, it is possible to alter this expectation\footnote{For a review, see~\cite{Dror:2020zru}.}. Examples include UV models with large charges~\cite{Agrawal:2017cmd,Agrawal:2018mkd}, clockworking~\cite{Agrawal:2017cmd,Agrawal:2018mkd,Choi:2015fiu,Farina:2016tgd,Coy:2017yex,Marques-Tavares:2018cwm}, kinetic mixing~\cite{Agrawal:2017cmd,Agrawal:2018mkd,Babu:1994id,Cicoli:2012sz,Higaki:2014qua,Bachlechner:2014hsa,Shiu:2015uva,Shiu:2015xda,Agrawal:2017eqm,Daido:2018dmu}, and discrete symmetries~\cite{Hook:2018jle,DiLuzio:2021pxd,DiLuzio:2021gos}. All of these represent distinct possibilities for $c_{a\gamma}^0$. Similar constructions can be performed for other axion couplings.  We note however that it seems that most, if not all, such constructions are inherently limited -- any attempt to arbitrarily boost/suppress axion couplings eventually encounters an obstacle.  For example, introduction of arbitrarily large couplings is prevented by the lowering of the QED Landau pole.\\

%==============================================================================================================================

%==============================================================================================================================
\subsection*{Stringy Models of the QCD Axion}
%==============================================================================================================================
Above we described the UV data defining a QCD axion model in the form of some QFT construction. Such a construction can be either embedded in, or replaced by, a string theory model. For our purposes, we will consider a 4d string model as being defined by a 10d string theory with 6 of the dimensions describe by a compact manfiold $X_6$, usually taken to be a Calabi-Yau manifold. Axions in such constructions arise from dimensional reduction of higher-form gauge potentials in the 10d theory, such as the 4-form $C_4$ in Type IIB string theory or the universal 2-form $B_2$ found in every critical superstring theory. However, axions can also arise from open string sectors associated with Dp-branes or as extremely light Kaluza-Klein modes in cases where $X_6$ contains highly warped regions~\cite{Hebecker:2015tzo,Hebecker:2018yxs,Carta:2021uwv}.

The number of axions one can expect in a string model depends on various details of the compactification, such as the Betti numbers counting particular sub-manifolds. One could expect dozens upwards of hundreds of axions arising in a string model, giving rise to the concept of the String Axiverse~\cite{Arvanitaki:2009fg,Cicoli:2012sz}. 

Within this landscape of axions, it is conceivable that there will be a QCD axion~\cite{Gaillard:2005gj,Svrcek:2006yi,Fox:2004kb}. The parameters of the QCD axion model will then be encoded in the details of the string compactification. For example, if the QCD axion arises as as zero mode of the $C_4$ potential, then the axion decay constant is set by the vacuum expectation value of scalar fields that control the size of sub-manifolds of $X_6$. Couplings, such as the QCD axion-photon interaction, are determined by other data such as the wrapping number of Dp-branes, intersection numbers of $X_6$, and/or quantized worldvolume magnetization.

From this perspective, it is clear that measuring the UV parameters of an axion model is tantamount to determining features of $X_6$ and related string data. However, the data provided by the manifold can be translated  into the UV data presented in the previous section.
Therefore, for our purposes, we can focus solely on $\datuv$ and be agnostic as to whether it arises from QFT or string theory. On the other hand, it is certainly possible that string models of the QCD axion have values for the elements of $\datuv$ that differ from the naive QFT perspective. For example, many string-based feature an axion decay constant on the order of $f_a \simeq 10^{16}$ GeV.

%%%%%%%%%%%%%%%%%%%%%%%%%%%%%%%%%%%%%%%%%%%%%%%%%%%%%%%%%%%%%%%%%%%%%%%%%%%%%%%%%%%%%%%%%%%%%%%%%%%%%%%%%%%%%%%%%%%%%%%%%%%%%%%
\section{Measuring the IR Data}
\label{sec:modeldisc}
In this section, we review the methods by which one could determine enough information in $\datir$ to reconstruct $\datuv$. In particular, we consider the operations of standard axion experiments and determine minimal networks to determine $\datuv$. In the subsequent section we apply these considerations to existing and proposed axion searches.

%==============================================================================================================================
\subsection{Haloscopes}
As their name suggests, haloscope experiments utilize SM couplings to probe dark matter axions in the Milky Way halo. Since the axion signal originates from the local dark matter distribution, the signal is proportional to $g_{SM}^2\rho_{DM}^{local}$. Furthermore, in resonance experiments, parameters are varied to scan frequencies until the axion mass is determined. Thus a resonant haloscope experiment can yield information on several elements of $\datir$ -- namely, $\rho_{DM}$, $g_{a-}$, and $m_a$. However we must keep in mind that generally the information on the first two parameters is degenerate and separate measurements must be made to fully disentangle the information made in a haloscope experiment.

We now discuss the operation of haloscopes in detail to more  fully describe the IR information they can probe. Photo-coupling haloscopes are widely discussed in the literature, so we will opt to omit a detailed discussion on these experiments. However, see ~\cite{Lee:2022mnc} for a thorough overview. We will simply emphasize that a signal at any of the photon-coupling haloscopes  yields a measurement of $\gagg^2\rho_{DM}^{local}$.

For haloscopes that probe fermion couplings, the relevant operator arises from the nonrelativistic limit of the first term in Eq. (\ref{eq:axioncoup1}). This gives the Hamiltonian
\begin{equation}
    H_{a\Bar{f}f}=-g_{af}\left(\nabla a\cdot \boldsymbol{\sigma}_f+\partial_t a\frac{\mathbf{p}\cdot\boldsymbol{\sigma}_f}{m_f}\right),
    \label{eq:axionfermionH}
\end{equation}
where $\boldsymbol{\sigma}_f$ is the fermion spin operator and $g_{af}$ is defined as in~\cref{eq:IRcoup}.
The first term in~\cref{eq:axionfermionH} is the so-called ``axion wind" and is probed by electron and nucleon haloscopes. We note that the axion gradient appears with the same spin coupling as an external magnetic field, so we shall refer to the combination $g_{af}\nabla a$ as a ``pseudo-magnetic field". The second term in~\cref{eq:axionfermionH} is the axioelectric term and is subdominant to the first term for the axion masses we will consider.  \\

\textbf{Electron Coupling:}
The existing proposals to probe electron couplings via haloscopes involve i) magnon experiments~\cite{Chigusa:2020gfs,Mitridate:2020kly}, ii) atomic transitions~\cite{Sikivie:2014lha}, and iii) utilizing the axion wind pseudo-magnetic field~\cite{Berlin:2023ubt}.

In the magnon solid state proposals~\cite{Chigusa:2020gfs,Mitridate:2020kly}, dark matter axions excite magnons (electron-spin waves) through the first term in~\cref{eq:axionfermionH}.
The axion mass range relevant for these experiments is $m_a\in[10^{-3},10^{-1}]$, which overlaps with the IAXO heliocope experiment discussed below. Axions are only able to couple to gapless magnons, but it is necessary that the magnon be gaped in order for an axion to be absorbed. There are a few ways this can happen. The first is by applying an external magnetic field to the sample to generate a gap as described in ~\cite{Chigusa:2020gfs}. Assuming $\rho_{DM}\sim 0.4$ GeV/cm$^3$, the current sensitivity projections reach the QCD axion line for $m_a\sim 10^{-3.8}$ eV.  Lower masses can in principle also be probed using this method, however, the emitted photons have very low energy and are not able to be detected using current techniques. A second method to gap magnons involves the use of materials with anisotropic interactions or nondegenerate Landé g-factors~\cite{Mitridate:2020kly}. Unlike with the external magnetic field method, the targeted axion mass is set by the material's resonances. So while the projected sensitivities reach the QCD axion line for some values of $m_a\in[10^{-3},10^{-1}]$ eV, the peaks are narrow and cannot be scanned through.

Another use for the gradient interaction in~\cref{eq:axionfermionH} lies in atomic transitions. As studied in~\cite{Sikivie:2014lha}, an atom can start in its ground state and be excited to some state $\ket{i}$ through the absorption of an axion via~\cref{eq:axionfermionH}. A laser is tuned on the atoms such that it excites atoms in the state $\ket{i}$ to some other excited state, but does not cause transitions from the ground state to other states. The atom then de-excites from this excited state to the ground state by releasing a photon, and this photon is counted. In general this signal will contain a part dependent on $g_{ae}$ and $g_{aN}$, as the axion can also interact with the spin of the nucleus. Signals due to each interaction can be isolated by using a variety of target atoms and by exploiting the fact that there are two or three transitions per target atom. This proposal looks at the mass range $10^{-6}-10^{-4}$eV. Taking $\rho_{DM}\sim 1$ GeV/cm$^3$ projections reach electron and nucleon couplings of $10^{-11}\text{ GeV}^{-1}$.

Finally, we consider the recent proposals to modify certain photon haloscope setups to also probe $g_{ae}$ ~\cite{Berlin:2023ubt}. The pseudo-magnetic field sourced by the axion wind which can be felt by electrons in a material and produce a magnetization current
\begin{equation}
\mathbf{J}_a^M=\nabla\cross\left((1-\mu^{-1})\mathbf{B}_{eff}\right).
\end{equation}
Here $\mu$ is the material's magnetic permeability. This current can be detected in multilayer setups with a signal again proportional to $g_{ae}\sqrt{\rho_{DM}}$. Current projections are for the mass range $10^{-6}-10^{-3}$ eV with better reach for lower masses. Taking the $\rho_{DM}=0.4\text{ GeV/cm}^3$ projections for $g_{ae}$ are $10^{-15}\text{ GeV}^{-1}$ at the low end of the mass range and $10^{-12}\text{ GeV}^{-1}$ at the high end.

\textbf{Nuclear Couplings:} Haloscopes probing the nucleon coupling of the QCD axion utilize either the first term in~\cref{eq:axioncoup1} or~\cref{eq:axionEDM}. The former induces a magnetic dipole moment, while the later induces an electric dipole moment. Both effects can be probed via nuclear magnetic resonance~\cite{Graham:2013gfa,Budker:2013hfa}, but we will consider the former type of experiment in this section. Current NMR experiments look for axions with $m_a\sim 10^{-12}$ eV$-10^{-6}$ eV. The operation of these NMR experiments hinges on the fact that the first term in~\cref{eq:axioncoup1} produces a pseudo-magnetic field felt by the nuclei in some sample material. We discussed a similar effect for electron couplings in the previous subsection, but the situation is more subtle here. The electron is a fundamental particle as far as we know, whereas the nucleus of an atom is a composite structure of protons and neutrons. Since an axion can couple differently to each of these baryons, one must take care in blindly applying~\cref{eq:axioncoup1} to nuclei.

Based on~\cref{eq:axioncoup1}, we shall write the axion-nuclei interaction Hamiltonian as~\cite{JacksonKimball:2017elr} 
\begin{equation}
    H_{aNN}=g_{aN}^{eff}\nabla a\cdot \mathbf{I},
    \label{eq:nucleonaxionH}
\end{equation}
where $\mathbf{I}$ is the nuclear spin and $g_{aN}^{eff}$ is some combination of $g_{ap}$ and $g_{an}$. 

To determine $g_{aN}^{eff}$, we can draw analogy with how the magnetic moment of a nucleus is calculated. The interaction of a nucleus with a real magnetic field is encapsulated in the Hamiltonian $H = -\vec{\mu}\cdot\vec{B}$, where the nuclear magnetic dipole moment (in units of the nuclear magneton) is calculated via the 
expectation value of the constituent nucleon spins:
\begin{equation}
    \mu=g_p\langle s_p^z\rangle +g_n\langle s_n^z\rangle+\langle l_p^z\rangle.
\end{equation}
The last term in this expression is the expectation value of the total
proton orbital angular momentum. In direct analogy with this, we can use the individual proton and neutron axion couplings in~\cref{eq:axioncoup1} to define an effective ``magnetic moment" for the axion interaction as 
\begin{equation}
    \begin{aligned}
            \mathbf{\mu}_a &\equiv g_{aN}^{eff}\mathbf{I}\\
            &= g_{ap}\langle s_p^z\rangle+g_{an}\langle s_n^z\rangle.
    \end{aligned}
\end{equation}
So that the effective nucleon coupling is
\begin{equation}\label{effective nuclear coupling}
    g_{aN}\equiv g_{aN}^{eff}=\frac{g_{ap}\langle s_p^z\rangle+g_{an}\langle s_n^z\rangle}{I}.
\end{equation}
The values of $\langle s_p^z\rangle$ and $\langle s_n^z\rangle $ come from nuclear physics and depend on number of nucleons. Values for a few nuclear models are given in ~\cite{Stadnik:2014xja}, but in general $g_{aN}^{eff}$ will be dominated by whichever nucleons are unpaired. More of the current NMR type experiments use nuclei with valence neutrons, like Xe$^{129}$ ~\cite{JacksonKimball:2017elr} or He$^3$ ~\cite{Chigusa:2023szl}, so it seems we are more likely to get a measurement of $g_{an}$ than $g_{ap}$ in the near future.

Beyond NMR, the interaction~\cref{eq:nucleonaxionH} can excite nuclear magnons~\cite{Chigusa:2023szl,Chigusa:2023hmz}. In the presence of an external magnetic field, superfluid $^3$He has a ferromagnetic phase, and the production of nuclear magnons in this phase is resonantly enhanced when the axion mass matches the Larmor frequency~\cite{Chigusa:2023szl}. Since the nuclear spin of $^3$He is determined by the neutron, this proposal is sensitive to the axion-neutron coupling 
$g_{an}$ around a mass of $m_a\sim 10^{-6}$ eV. In the mass range $m_a\sim 10^{-6}\text{ eV}-10^{-4}$ eV, another proposal suggests utilizing a magnet with a strong hyperfine interaction to detect dark matter axions~\cite{Chigusa:2023hmz}. The proposal sample in this case is MnCO$_3$, which in contrast to $^3$He, the nuclear spin arises primarily from the proton, so for this experiment $g_{aN}^{eff}\approx g_{ap}$. Interestingly, axions can also excite the electrons in MnCO$_3$. These electrons can then excite the nuclear spins, giving another channel for axion detection sensitive to $g_{ae}$.

Lastly, the nuclear couplings may also be measured using atomic transitions ~\cite{Sikivie:2014lha}. This is described in more detail in the previous section. However, there are currently no experiments attempting to measure $g_{aN}$ in this way. As mentioned in the last section, this proposal is for axion masses of $10^{-6}-10^{-4}$ and can reach nucleon couplings down to $10^{-11} \text{ GeV}^{-1}$ assuming $\rho_{DM}\sim 1$ GeV/cm${^3}$.\\

\textbf{EDM}:
In the presence of dark matter, the interaction~\cref{eq:axionEDM} induces an oscillating nuclear EDM. In the non-relativistic limit, we obtain the effective Hamiltonian~\cite{Garcon_2017,JacksonKimball:2017elr}
\begin{equation}
    H_{EDM}=-\mathbf{d}_n(t)\cdot\mathbf{E},
    \label{eq:EDMHamiltonian}
\end{equation}
where $\mathbf{E}$ is an applied electric electric field and
\begin{equation}
   \mathbf{d}_n(t)=g_d a_0\cos{(\omega_a t)}\frac{\mathbf{I}}{I}. 
\end{equation}
NMR type experiments, similar to those described in the last section, can be used. At the resonance condition, the axion-induced EDM oscillates at the Larmor frequency and the NMR signal is proportional to $g_d\sqrt{\rho_{DM}}$.

A separate proposal to probe the EDM operator utilizes dielectrics~\cite{PhysRevD.42.1847,Berlin:2022mia}.
In dielectrics, spin-polarized atoms have an axion-induced EDM leading to a polarization density proportional to $\mathbf{d}_n(t)$,which can be measured via the induced current. Recently it has been shown that sensitivity in the range $m_a\sim 10^{-6}-10^{-4}$ eV range can be reached for the proper material choice ~\cite{Berlin:2022mia}. The signal in these ``polarization haloscopes" is again proportional to $g_d\sqrt{\rho_{DM}}$.
%==============================================================================================================================

%==============================================================================================================================
\subsection{Helioscopes}
We now consider which elements of $\datir$ can be probed by helioscope experiments. In contrast to haloscopes, helioscopes do not probe the local dark matter density, but instead axions that arrive on Earth after being produced in the sun. These axions are then converted into photons in an apparatus and detected. The conversion probability of an axion with mass $m_a$ and energy $E_a$ traveling a distance $L$ along a magnetic field $B$ is 
\begin{align}
    P_{a\rightarrow\gamma} = \bigg(\frac{g_{a\gamma}B}{q}\bigg)^2\sin^2\bigg(\frac{qL}{2}\bigg)
\label{eq:helioprob}
\end{align}
where $q=\frac{m_a^2}{2E_a}$, and we have assumed a vacuum ~\cite{CAST:2008ixs}. Within the sun, there are several axion production mechanisms occurring: Primakoff conversion, Compton scattering, bremsstrahlung, and atomic recombination and de-excitation. Primakoff conversion involves only the axion-photon coupling and the differential axion flux is proportional to $\gagg^2$. The remaining three processes are proportional to $\gae^2$ ~\cite{Redondo:2013wwa}. Combining these scalings with~\cref{eq:helioprob}, we see that the helioscope signal goes as $\gagg^4$ for Primakoff conversion and $\gagg^2\gae^2$ for atomic recombination \& deexcitation, bremsstrahlung, and Compton scattering.

In this general situation, one might be concerned that the photon and electron processes make measurements of $\gagg$ or $\gae$ difficult. Luckily, this is not necessarily the case.  As argued in~\cite{Jaeckel:2018mbn}, it is possible to simultaneously measure $\gagg$ and $g_{ae}$ when the fluxes from the Primakoff and electron processes are comparable. In cases where the Primakoff process dominates, $\gagg$ can be measured and an upper bound can be placed on $g_{ae}$; if the eletron processes dominate, $g_{a\gamma}g_{ae}$ can be measured, and an upper bound can be placed on $\frac{g_{a\gamma}}{g_{ae}}$.

It may also be possible to get a measurement of the nucleon coupling from a helioscope as described in ~\cite{DiLuzio:2021qct}. The dominant nuclear process in the sun which produces axions is the de-excitation of the first excited state of $^{57}$Fe. This process emits axions with an energy of 14.4 keV. The signal is proportional to $g_{a\gamma}^2(g_{aN}^{eff})^2$, where $g_{aN}^{eff}$ is the effective nuclear coupling for $^{57}$Fe.~\cref{fig:solarfluxes} shows that, for sensible choices of couplings, fluxes from the photon, electron, and iron processes are visible and have distinct peaks.

Helioscopes also have some ability to measure $m_a$ from the mass dependence in Eq. (\ref{eq:helioprob}). The helioscope signal is coherent for $qL<\pi$, but if this condition is only mildly violated, the oscillations due to the $\sin^2$ term are strong enough to be measured, but do not destroy the signal entirely ~\cite{Dafni:2018tvj}. For lighter axions these oscillations can be seen at lower energies. The coherence condition can be extended to higher masses by filling the detector with a buffer gas. The photons acquire an effective mass of $m_\gamma=\sqrt{4\pi\alpha_{EM}n_e/m_e}$, where $n_e$ is the electron density in the gas. The conversion probability is modified to
\begin{align}
\begin{split}
    P_{a\rightarrow \gamma}=&\left(\frac{g_{a\gamma} B}{2}\right)^2\frac{1}{q^2+\Gamma^2/4}\\
    &\times\left(1+e^{-\Gamma L}-2e^{-\Gamma L/2}\cos{(qL)}\right). 
    \label{eq:helioprobgas}
\end{split}    
\end{align}
Where $\Gamma$ is the inverse photon absorption length and the momentum transferred is now $q=|m_a^2-m_\gamma^2|/2E_a$. A range of axion masses can be searched for by modifying the density of the gas.

We also note the discussions on discriminating between a massless and massive axion signal in~\cite{Dafni:2018tvj} as well as the ability for an axion signal from IAXO to inform solar models~\cite{Hoof:2021mld} and measure the Sun's magnetic field~\cite{OHare:2020wum}

Finally, we mention another prospect for measuring $g_{aN}$, through terrestrial iron experiments ~\cite{Moriyama:1995bz,Moriyama:1998tx,Krcmar:1998xn} The advantage of this type of experiment is that the signal would only depend on $g_{aN}$ and not any other axion coupling. However, a quick calculation using the formulas in ~\cite{Moriyama:1995bz} estimates that at least $10^6$ kg of iron-57 would be needed to potentially get a signal, so currently this is not a realistic experimental prospect.\\
\begin{figure}
\centering
\includegraphics[width=\columnwidth]{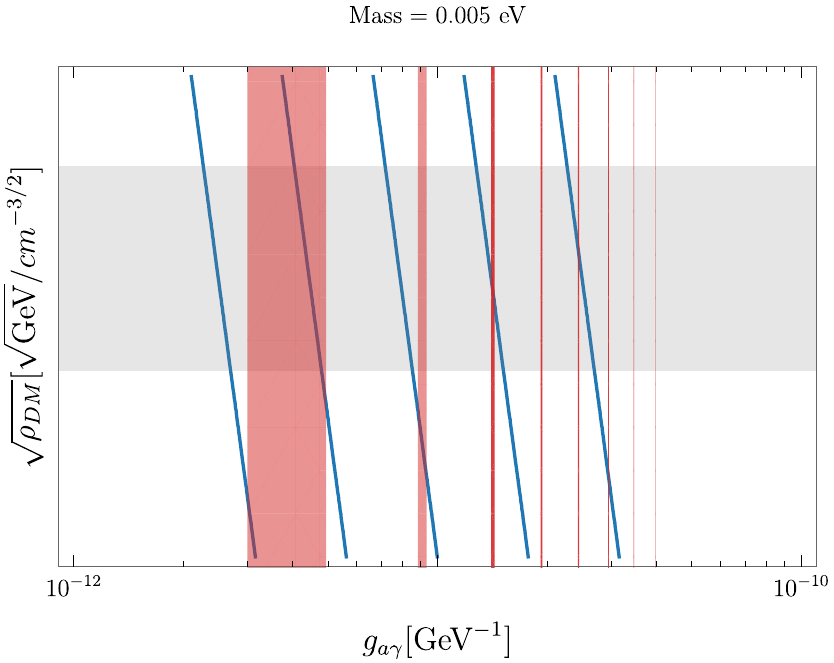}
\caption{Haloscope \& Helioscope Combined Data. The the local dark matter density window is taken as $\rho_{DM}^{local}= 0.39\pm 0.03$ GeV$\cdot$ cm$^{-3}$~\cite{Read:2014qva,Green:2017odb}. If the parameters are favorable, this combination will provide the best measurement of the axion density in the halo.} 
\label{fig:DM_uncertainty}
\end{figure}

%==============================================================================================================================

%==============================================================================================================================
\subsection{Astrophysical Observations}
Similar in spirit to helioscopes, there are also proposals to detect axions produced through electron bremsstrahlung in white dwarfs~\cite{Dessert:2021bkv}. These axions can convert to photons in the white dwarf magnetoshpere, which can then be detected on earth. A positive signal would provide us a measurement of the product $g_{ae}g_{a\gamma}$. This would be extremely interesting, as a positive signal here, paired with an independent measurement of $\gagg$, would allow us to 
obtain a value for the electron coupling. The current bounds are strongest for $m_a\lesssim 10^{-5}$ eV, ruling out $g_{a\gamma}g_{ae}\gtrsim6.6\times10^{-29}\text{ GeV}^{-2}$.

A relative of the above white dwarf production can be seen in neutron stars. Axions can be produced in the magentosphere of a neutron star via the Chern-Simons coupling to the photon. For $10^{-9}\lesssim$ eV $m_a\lesssim 10^{-4}$ eV, a significant portion of axions are gravitationally bound to the neutron star, forming an ``axion cloud"~\cite{Noordhuis:2023wid}. For $m_a\gtrsim 10^{-7}$ eV, the axion cloud primarily radiates energy through axions converting into photons. These photons produced multiple distinct signatures, including one at about $m_a$, which can in principle be detected. Interestingly, the signal depends only on a single IR parameter, $\gagg$.

%==============================================================================================================================

%==============================================================================================================================
\subsection{Other Experiments}
Here we briefly describe the idea behind table-top experiment proposed in ~\cite{Batllori:2023gwy}, called  WISP Searches
on a Fiber Interferometer (WISPFI). A laser is sent through an interferometer, which has one arm placed in a magnetic field. Photons passing through the magnetic field can convert to axions via the Primakoff effect with the probability of conversion $P_{\gamma\rightarrow a}\propto g_{a\gamma}^2$. {The amplitude of this beam can be compared to the one that did not pass through the field, with the amplitude difference proportional to $g_{a\gamma}^2$. The current proposal is for $m_a\in [28 \text{ meV},100 \text{ meV}]$, but there is potential to explore a larger range. 
%==============================================================================================================================

%==============================================================================================================================
\subsection{Extracting the Data}

In the previous subsections, we described various axion experiments and which elements of $\datir$ they probe. We now consider how to use all of this technology. We will discuss idealized scenarios in which one or more of the above experiments find a positive signal and thereby provide values for elements of $\datir$ (or combinations thereof). In the subsequent section we will determine which, if any  of these situations can arise.\\
\begin{comment}
\textbf{Scenario I:}
\begin{itemize}
    \item Helioscope: measures $\gagg$, $\gann$, $\gae$, and $m_a$ 
    \item Haloscope (any): measures $g_{SM}^2\rho_{DM}^{local}$ and $m_a$ 
\end{itemize}
In this scenario, a helioscope can produce a wealth of data in $\datir$.

\textbf{Scenario II:}\\
\begin{itemize}
    \item Heliscope: sees signal but cannot resolve $\gagg$ and $g_{ae}$, can still measure $m_a$
    \item Astrophysical observations measure: $g_{a\gamma}g_{ae}$
    \item Haloscope (photon or electron): measures $g_{a\gamma}^2\rho_{DM}$ or $g_{ae}^2\rho_{DM}$ and $m_a$ 
    \item Haloscope (nuclear): measures $g_{aN}^2\rho_{DM}$ and $m_a$
\end{itemize}

 \textbf{Scenario III:}\\
 \begin{itemize}
    \item Haloscopes (all): distinct haloscope experiments provide $g_{SM}^2\rho_{DM}$ and $m_a$
    \item EDM measurement
\end{itemize}

\textbf{EDM Measurement:}\\
\begin{itemize}
    \item $g_d\sqrt{\rho_{DM}}$ and $m_a$
    \item $g_d\propto f_a^{-1}\propto m_a$ $\rightarrow$ $g_d$ determined by $m_a$
    \item Can extract $\rho_{DM}$ from one experiment
\end{itemize}

 \textbf{Scenario IV:}\\
\begin{itemize}
    \item Haloscope (photon): measures $\gagg^2\rho_{DM}^{local}$ and $m_a$ 
    \item Haloscope (nuclear): measures $\gann^2\rho_{DM}^{local}$ and $m_a$ 
    \item Astrophysical observations: WD measure $\gae\gagg$ 
\end{itemize}
\end{comment}

\textbf{Scenario I:}
Let us start by disregarding helioscopes and consider a situation where a haloscopes provide one or more positive signals. Due to the prevalence of photon-coupling haloscopes, it is natural to consider the signal as arising from one of these experiments. In this case, we would obtain a measurement of $g_{a\gamma\gamma}^2\rho_{DM}$ and $m_a$. Other haloscope measurements can be performed to find $g_{ae}^2\rho_{DM}$ and $g_{aN}^2\rho_{DM}$. All of these measurements depend on the dark matter density, so we would be able to deduce ratios of the couplings (i.e. $g_{a\gamma\gamma}/g_{ae}$), but we cannot resolve any coupling individually. However, if a measurement is made in an EDM haloscope, we actually get more information than it first seems. While an EDM haloscope measures the product $g_d\sqrt{\rho_{DM}}$, $g_d$ is fully determined by $m_a$ for a QCD axion. Thus once $m_a$ is determined, an EDM signal provides a value for the dark local matter density. Pairing this with other haloscope signals, one can potentially determine each of the photon, nucleon, and electron couplings.\\

\textbf{Scenario IIa}
We next consider the idealized situation of the previous section where a helioscope can determine the values of $\gagg$, $\gae$, and $g_{aN}$, and potentially an estimate of the axion's mass. If a positive signal can then be obtained from a haloscope, the helioscope coupling data allows for a determination of $\rho_{DM}$. This is perhaps the most fortuitous scenario in that we get five parameters from only two experiments. However, $g_{aN}=g_{aN}^{eff}$ is not a defining parameter of the IR theory; it is a linear combination of $g_{an}$ and $g_{ap}$. Another nuclear coupling measurement is needed to individually resolve the proton and neutron coupling, most likely from another haloscope.\\

\textbf{Scenario IIb:}
A variation to the above situation is that we obtain values for $\{\gagg,\gae, m_a\}$ from a helioscope but not $g_{aN}$.
We would then need a supplemental haloscope measurement of $g_{aN}^2\rho_{DM}$ to get the nuclear coupling. We would then need additional work to disentangle the distinct proton and neutron couplings.\\

\textbf{Scenario III:}
Alternatively, we could have a situation where a positive helioscope signal is in a regime where the Primakoff flux dominates. Here, we expect to extract the value of $\gagg$ and at most put an upper bound on $g_{ae}$. If there is a signal from the iron-based proposal above, we can measure $g_{aN}$, or, if no signal is seen, an upper bound can be placed. Similar to the last scenario, a supplemental photon haloscope signal would give $m_a$ and $\rho_{DM}$. The other couplings can then be searched for in haloscopes, though this may be difficult.\\

\textbf{Scenario IV}
Finally, we consider the case with an electron-dominated helioscope measurement. Here, we do not measure any coupling directly, but only the product $\gagg\gae$. Further information would be required from other experiments. A haloscope signal would only yield a measurement of the product $g\sqrt{\rho_{DM}}$, for one of the couplings $g$, and $m_a$. Therefore, unlike the previous scenarios, we would be unable to extract any individual coupling at this point. If $m_a$ falls in the range of a WISPFI-type experiment, we could obtain a direct measurement of $\gagg$. Alternatively, if we can measure $g_{ae}\sqrt{\rho_{DM}}$ at a electron-coupling haloscope, this can be combined with the results of the helioscope and photon haloscope to determine $\gagg$, $g_{ae}$, and $\sqrt{\rho_{DM}}$. The nucleon couplings would all have to be determined independently.

%==============================================================================================================================

%%%%%%%%%%%%%%%%%%%%%%%%%%%%%%%%%%%%%%%%%%%%%%%%%%%%%%%%%%%%%%%%%%%%%%%%%%%%%%%%%%%%%%%%%%%%%%%%%%%%%%%%%%%%%%%%%%%%%%%%%%%%%%%

%%%%%%%%%%%%%%%%%%%%%%%%%%%%%%%%%%%%%%%%%%%%%%%%%%%%%%%%%%%%%%%%%%%%%%%%%%%%%%%%%%%%%%%%%%%%%%%%%%%%%%%%%%%%%%%%%%%%%%%%%%%%%%%
\section{Experimental Prospects}
\label{sec:axpro}
In the previous section, we determined how various experimental setups probe different subsets of $\datir$. We now apply this knowledge to the concrete axion experimental program in order to determine to what extent we will be able to i) determine $\datir$ and ii) utilize this to determine $\datuv$.

As axion experiments span a wide range in the mass and coupling parameter space, our discussion will proceed by considering disjoint mass ranges and describe potential discovery avenues via the scenarios listed at the end of~\cref{sec:modeldisc}. We will largely consider only the mass range
\begin{equation}
10^{-12}\text{ eV}\lesssim m_a\lesssim 10 \text{ eV}
\label{eq:axionmasses}
\end{equation}
%==============================================================================================================================
\subsection{Mass Range I: The Low Mass Case}
We define a low mass range as the interval 
\begin{equation}
    10^{-12}\text{ eV}\lesssim m_a\lesssim 10^{-6} \text{ eV}.
\label{eq:lowmass}
\end{equation}
Photon-coupling haloscope experiment proposals in this region, including DM-Radio~\cite{Chaudhuri:2014dla}, SRF-m$^3$~\cite{Berlin:2020vrk}, WISPLC~\cite{Zhang:2021bpa}, aLIGO~\cite{Nagano:2019rbw}, DANCE~\cite{Michimura:2019qxr}, and ADBC~\cite{Liu:2018icu}. Of these, only DM-Radio and SRF-m$^3$ are able to probe the typical QCD axion parameter space.

There are also proposals for low-mass nucleon-coupling haloscopes in the form of CASPEr-Wind~\cite{Graham:2013gfa,Budker:2013hfa}. These utilize Helium and Xenon samples and thus probe the axion-neutron interaction.  However, these do not reach the typical QCD axion parameter space. Currently, there are no proposals probing the axion-proton interaction. The only proposal to measure the electron coupling of axions in this mass range is through WD observations ~\cite{Dessert:2021bkv}. However, the projections likewise do not reach the usual QCD axion parameter space. 

Also operating within the mass range in~\cref{eq:lowmass} is CASPEr-Electric~\cite{Budker:2013hfa} and other EDM experimental probes~\cite{Berlin:2022mia}. Such proposals are particularly important, since an EDM coupling is necessary to definitely prove an axion is \textit{the} QCD axion. Unfortunately, the proposed search strategy of CASPEr-electric in probes the QCD axion line only for masses $m_a\lesssim 10^{-9}$ eV. However, the results of~\cite{Dror:2022xpi} updates and improves the original CASPEr projections. This opens the opportunity for reaching the QCD axion for higher masses by increasing the time spend on a given mass point. This is displayed by the solid contours in~\cref{fig:EDMprobe} for Phase 1 and Phase 2 of~\cite{Budker:2013hfa}. Phase 1 (Phase 2) is taken to have an applied magnetic field of strength $B= 10$ T ($B=20$ T) and a transverse spin-relaxation time $T_2 = 10^{-3}$ s ($T_2 =1$ s). However, compared to the original proposal, we have increased the integration time to $T= 10^5$ s. This allows for the Phase 2 CASPEr-electric projections to reach the QCD axion line. A similar comment applies to the proposal with $^3$He superfluid~\cite{Chigusa:2023szl} for even a gram-scale experiment.
\begin{figure}
    \centering
    \includegraphics[width=\columnwidth]{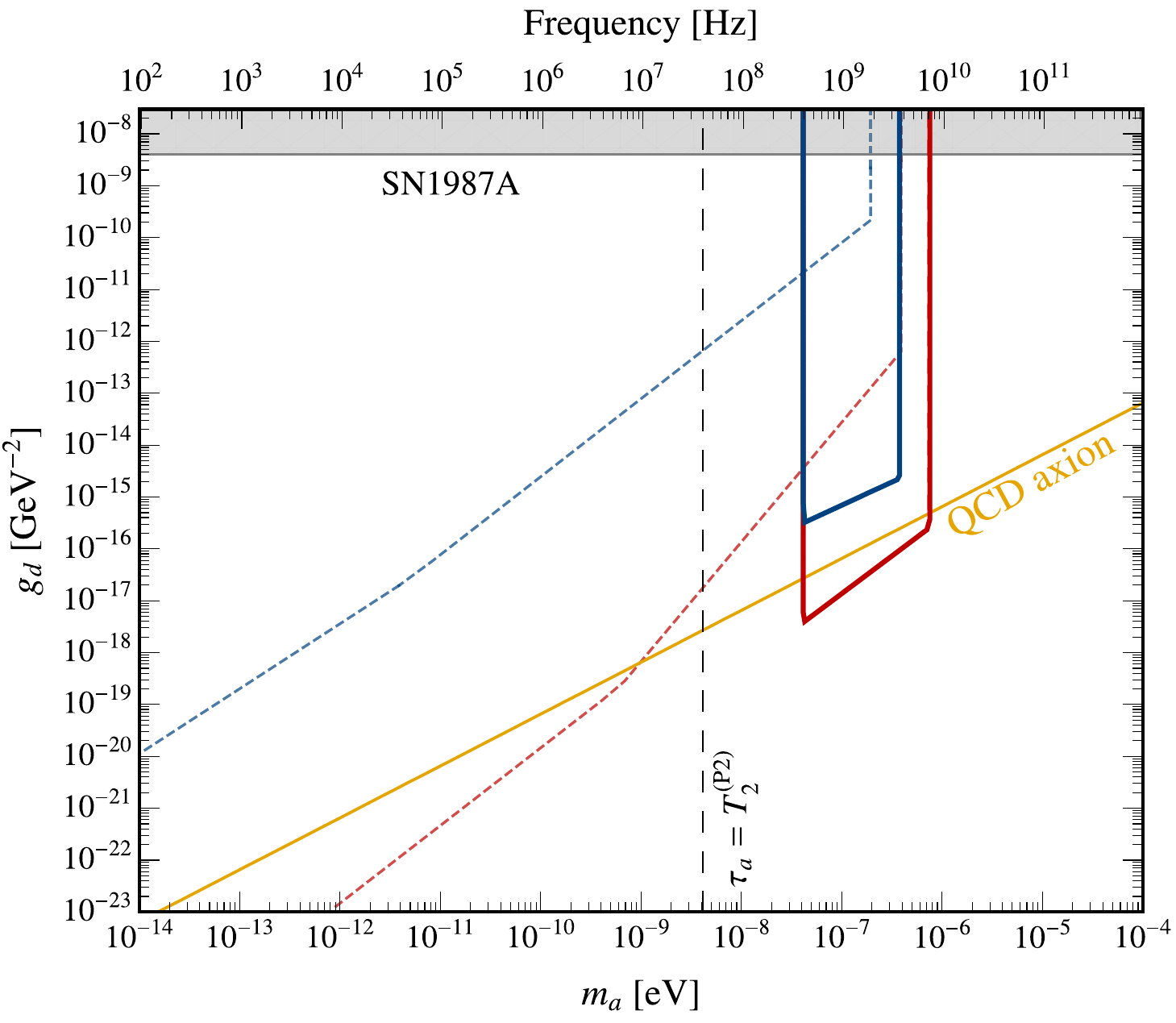}
    \caption{EDM coupling projections for CASPEr-electric 
    for Phase 1 (blue) and Phase 2 (red). Dashed lines correspond to the original projections of~\cite{Budker:2013hfa} while the solid curves utilize the results of~\cite{Dror:2022xpi} but with an integration time increased to $10^5$ s.}
    \label{fig:EDMprobe}
\end{figure}

Synthesizing these considerations with the scenarios of the previous section, a potential future situation in this mass range involves a positive signals at a photon and EDM haloscope. This is an incomplete version of \textbf{Scenario I} -- we would be able to determine the axion is indeed the QCD axion, as well as measure the $\datir$ elements $m_a$, $\gagg$, and $\rho_{DM}$. In terms of the $\datuv$, one would be able to extract $c^0_{a\gamma}$ and $f_a$. This may give us hints towards the underlying axion model, but as ~\cite{DiLuzio:2017pfr} shows $\frac{\emano}{\cano}$ can take on many values for both KSVZ and DFSZ axions. To further model discrimination, focus should be on developing experiments measuring axion-fermion couplings at the now known $m_a$. 

%==============================================================================================================================

%==============================================================================================================================
\subsection{Mass Range II: The Intermediate Mass Case}
We now turn to the intermediate mass range 
\begin{equation}
    10^{-6}\text{ eV} \lesssim m_a \lesssim 10^{-3}\text{ eV}
\label{eq:midmass}
\end{equation} 
Within this range, there is a wealth of photon-coupling haloscope proposals, including ADMX~\cite{Asztalos_2010,Du:2018uak,Braine:2019fqb,Boutan:2018uoc,Stern:2016bbw,Crisosto:2019fcj,ADMXwhite:2020}, ALPHA~\cite{Lawson:2019brd}, FLASH~\cite{FLASH2022}, QUAX~\cite{Alesini:2019ajt,Alesini:2020vny,Alesini:2022lnp,QUAX:2023gop,QUAX:2024fut,QUAX2021}, BabyIAXO-RADES~\cite{Ahyoune:2023gfw}, DALI~\cite{DeMiguel:2023nmz}, ORGAN~\cite{McAllister:2017lkb,Quiskamp:2022pks,Quiskamp:2023ehr,Quiskamp:2024oet},CAPP~\cite{Lee:2020cfj,Jeong:2020cwz,CAPP:2020utb,Lee:2022mnc,Yoon:2022gzp,Kim:2022hmg,Yi:2022fmn,Adair:2022rtw,Kim:2023vpo,Yang:2023yry,CAPP:2024dtx}, and MADMAX~\cite{Beurthey:2020yuq}. There are also tunable plasma haloscopes~\cite{Lawson:2019brd}, which have more potential to probe higher masses compared to other haloscopes. 

For axion-electron couplings, the magnon-based haloscopes~\cite{Chigusa:2020gfs} provide a
potential discovery channel for axions as their projections reach typical QCD values for DFSZ-type axions in the upper end of the mass range in~\cref{eq:midmass}. Another possibility could be astrophysical observations of white dwarves~\cite{Dessert:2021bkv}, which measure the product $\gagg\gae$. Unfortunately, the limits placed in~\cite{Dessert:2021bkv}  are far above the typical QCD axion window, but an improvement by an order of magnitude or so would allow this method to begin probing the standard DFSZ QCD axion parameter space.
The measurement, however, is subject to the uncertainty in the size of the magnetic field of a given white dwarf. Optimistic projections from axion-induced atomic transitions ~\cite{Sikivie:2014lha} reach the QCD band for part of this mass range, but there have been few attempts to devise experiments utilizing this method~\cite{Braggio:2017oyt, Vergados:2018qdb}. The best current projections come from axion-wind multilayer setups ~\cite{Berlin:2023ubt}, which reach well into the QCD band and cover nearly all of expected DFSZ values.

Nuclear magnon experiments~\cite{Chigusa:2023hmz, Chigusa:2023szl} measuring both $g_{an}$ and $g_{ap}$ operate in the range~\cref{eq:midmass}. The proposal in~\cite{Chigusa:2023hmz}
covers typical values for DFSZ and KSVZ type axions for $m_a\sim 10^{-6}$.

As for the EDM coupling, the mass range in~\cref{eq:midmass} is above the current projections for CASPEr-electric. However, ~\cite{Berlin:2022mia} provides a probe of the gluon coupling of the axion dark matter signal, and projections for upgraded polarization haloscopes cover predicted QCD values for most of this mass region. In terms of $\datuv$, we would be able to discern $f_a$, $c_{a\gamma}$, and $c_{ae}$.

Examining the above proposals, it appears that \textbf{Scenario I} would be the most likely situation for positive axion discovery in the mass range~\cref{eq:midmass}. One could imagine a positive signal from one of the plenitude of photon haloscope discussed above, yielding a measurement of $g_{a\gamma}\sqrt{\rho_{DM}}$ and $m_a$. With such a signal, it would be beneficial to expend the effort to measure $g_{d}\sqrt{\rho_{DM}}$ from a polarization haloscope. Then, as in the low mass case, we would be able to determine if the particle detected is indeed the QCD axion and resolve $\sqrt{\rho_{DM}}$ from the haloscope results. Since the prospects for other couplings are more robust here than in the low mass case, we can go further. Development of an electron-coupling haloscope following the proposal of~\cite{Berlin:2023ubt} would provide key insight in this region, as a positive signal would give a value of $\gae$. Furthermore, such a signal would rule out KSVZ axions, as their smaller electron couplings would not be visible to the proposals of~\cite{Berlin:2023ubt}. On the other hand, if no signal is seen, we can start to rule out parts of the DFSZ parameter space.

Unfortunately, for axion-nucleon couplings, there are limited prospects for the mass range~\cref{eq:midmass} and new proposals are needed. If the axion is known to be DFSZ type from the electron haloscope, measuring $g_{an}$ or $g_{ap}$ gives us $\beta$ (since $f_a$ is already known). This can be used to distinguish between DFSZ types 1 and 2. If no signal was seen in the electron haloscope, measuring the nucleon couplings will be helpful in determining if the axion is KSVZ or DFSZ type.
%==============================================================================================================================

%==============================================================================================================================
\subsection{Mass Range III: The High Mass Case}
%==============================================================================================================================
Finally, we consider the upper end of the mass range in~\cref{eq:axionmasses} -- namely the interval
\begin{equation}
    10^{-3}\text{ eV}\lesssim m_a\lesssim 10^1 \text{ eV}
\label{eq:highmass}
\end{equation}
The relevant experiments in this range are the helioscope IAXO~\cite{IAXO:2013len,Armengaud:2014gea,IAXO:2019mpb,IAXOwhite:2020} and its variants babyIAXO~\cite{Abeln:2020ywv}, IAXO+~\cite{IAXO:2019mpb}, and IAXO (SN)~\cite{Ge:2020zww} and the photon-coupling haloscopes LAMPOST~\cite{Baryakhtar:2018doz}, BREAD~\cite{BREAD:2021tpx}, EQC~\cite{Fan:2024mhm}, CADEx~\cite{Aja:2022csb}, and TOORAD~\cite{Schutte-Engel:2021bqm}.

It is important to note that part of this mass range is ruled out for usual QCD axion models. Constraints from the cooling of SN 1987 give $m_a< 16\text{ meV}$ for KSVZ axions and $m_a\lesssim 30 \text{ meV}$ for DFSZ axions ~\cite{Buschmann_2022}. If the mass of detected particle exceeds these constraints, could be an ALP, or need some non-trivial model building to explain.

Currently, there are no proposals probing axion-nucleon couplings or the EDM in the range~\cref{eq:highmass}. There is a possibility to probe the electron coupling through magnons ~\cite{Mitridate:2020kly}, but as described earlier, this is only possible for small ranges of $m_a$.

Given the dearth of experiments operating in this range, a positive signal in the mass range ~\cref{eq:highmass} would likely
lead to an incomplete version of \textbf{Scenario II, III,} or \textbf{IV}. We will take helioscopes to form the backbone of our discussion here and assume that a positive signal is found at IAXO or a similar experiment.

Let us first consider the optimistic case of \textbf{Scenario IIb} where the helioscope is able to extract $\gagg$ and $g_{ae}$, as discussed in the previous section, and an estimate of $m_a$. The strong electron flux signal required for this scenario indicates that if the particle is the QCD axion, it is of DFSZ type. To proceed further, we would need to focus efforts to detect the axion in a photon-coupling haloscope to extract $\gagg$ and $\rho_{DM}$. This would also give a better measurement of $m_a$. If we also measure $g_{aN}$ from the iron signal \cite{DiLuzio:2021qct}, we can use this in combination with $g_{ae}$ to differentiate between DFSZ types I and II. Since the nucleon and electron couplings depend on $f_a$ in the same way, in this case we would know the ratio $c_{aN}/c_{ae}$. For specific DFSZ models $c_{aN}$ is determined by $c_{au_i}$ and $c_{ad_i}$, so here we can put constraints on the ratios of a few  $\datuv$ parameters.

A slightly less fortuitous situation is \textbf{Scenario III}, where instead of measuring $g_{ae}$ from a helioscope, we can only put an upper bound on it. Interestingly, the addition of a $g_{aN}$ measurement can potentially rule out DFSZ models. An upper bound on $g_{ae}/g_{aN}$ puts constraints on $\datuv$ parameters $c_{ae}$, $c_{au_i}$, and $c_{ad_i}$ and rules out a range of the $\beta$ values in~\cref{eq:betas}. If the value of $g_{aN}$ measured fits both KSVZ and DFSZ models, the DFSZ model could be ruled out due to the constraint on $\beta$. Next, we would again look for a signal from a photon haloscope to obtain measurements of $m_a$ and $\rho_{DM}$. If the iron signal was not measured in IAXO, it would be useful to use a haloscope to measure the nucleon couplings. Future or improved experiments would be needed to do this for the masses considered here.

Finally, we can consider axion detection playing out through \textbf{Scenario IV}. If the IAXO signal is dominated by the electron flux, we measure the combination $g_{ae}g_{a\gamma}$ instead of an individual coupling. Once again, the large value of $g_{ae}$ required to generate the signal would rule out the KSVZ axions. As the scenarios above, it would be advantageous to supplement the helioscope signal with one from a photon-coupling haloscope. The prospects for this are slim as the lack of photon flux in the helioscope signal implies that  $\gagg$ is small. As described in \textbf{Scenario IV}, for $m_a\in [28 \text{ meV},100 \text{ meV}]$ WISPFI can measure $\gagg$ directly. This could then be used to extract the axion-electron coupling from the IAXO signal without involving a haloscope or the dark matter density. Otherwise, we can combine the results from a photon and electron haloscope to extract the coupling values. The prospects for electron haloscopes in this mass range cannot search for different values of $m_a$. But if the mass is known, it may be possible to target that specific mass. If the particle detected is indeed the QCD axion, combining $g_{ae}$ and $m_a$ gives the value of the $\datuv$ parameter $c_{ae}$. However, another experiment would need to confirm the particle as the QCD axion.

In the preceding scenarios, we outline how the value of $\rho_{DM}$ can be found by combining data from a helioscope and haloscope. Our ability to find the value of $\rho_{DM}$ in this way will depend on the uncertainty of $\gagg$ from the helioscope. Fig. \ref{fig:DM_uncertainty} shows what this looks like for different measurements of $g_{a\gamma}\sqrt{\rho_{DM}}$ and $\gagg$. Details of how the uncertainty of $\gagg$ are given in~\cref{app:Asimov}. For simplicity, we plot only the Primakoff dominated case, so we do not consider $g_{ae}$ or its uncertainty.

In the above cases we have assumed we have some ability to distinguish the photon and electron signal in IAXO. However, as outlined in ~\cite{Jaeckel:2018mbn}, this will not be true for the entire operating mass range, but only up to $m_a\sim0.2$ eV. Even for masses below this, it may be the case that $\gagg$ and $g_{ae}$ are small and cannot be individually resolved. Thus, we may have to rely on other experiments to measure couplings after a positive IAXO signal. WISPFI can help fill the gap, but current projections are sensitive to a small mass range. Photon haloscopes can be used to measure $g_{a\gamma}\sqrt{\rho_{DM}}$ and $m_a$. If we already know the photon coupling, we get $\rho_{DM}$. Otherwise, knowledge of $m_a$ will likely be useful in developing further experiments.

Finally we note that it is possible the axion falls in this mass range, but IAXO cannot see it. The projections do not cover the full range of QCD values for $m_a\sim 10^{-3}-10^{-2}$ eV. Such axions can still be searched for in the halsocope BREAD.
\begin{figure}
    \centering
    \includegraphics[width=\columnwidth]{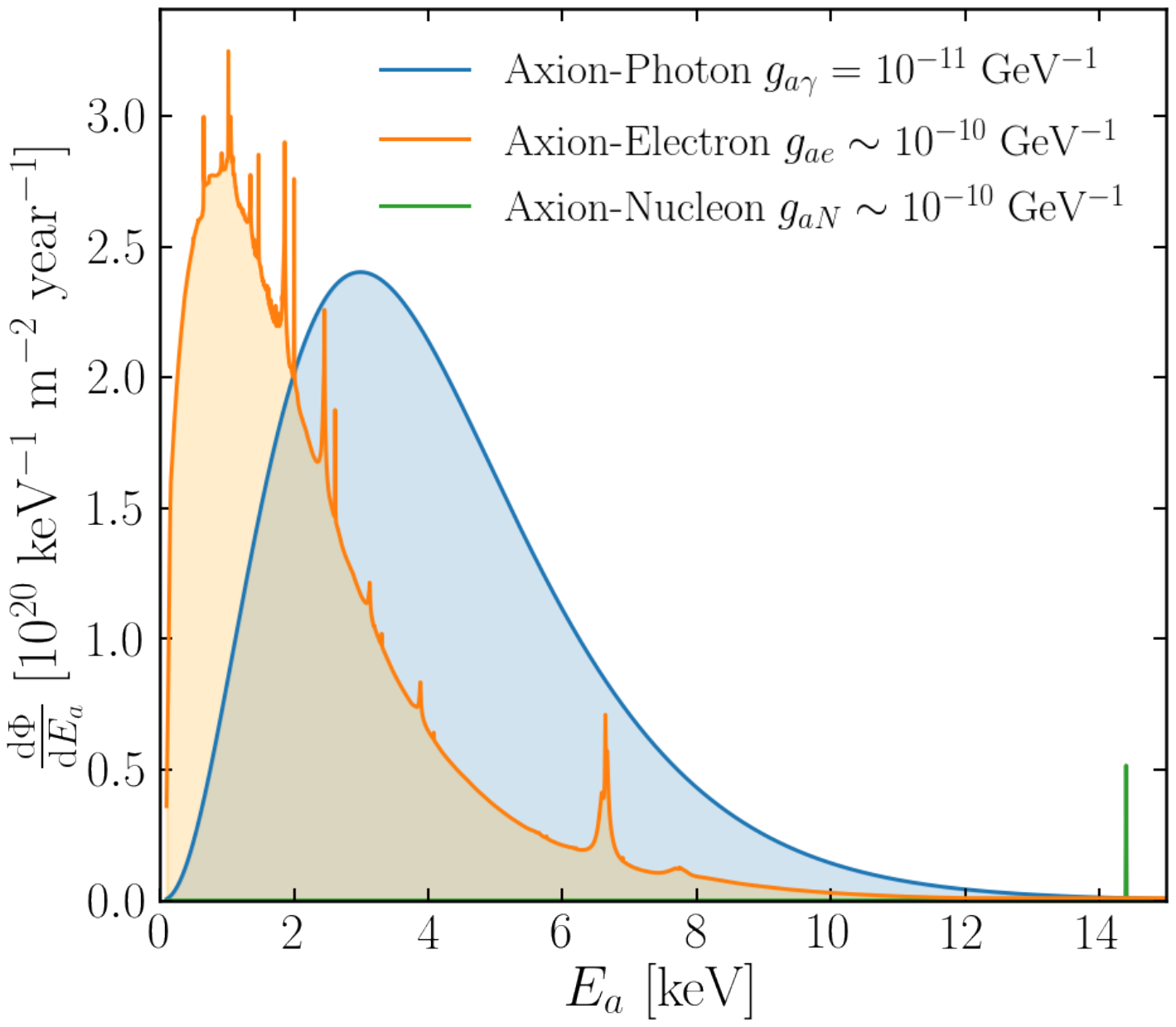}
    \caption{Expected solar axion fluxes in a helioscope from the photon, electron, and iron processes. The photon and electron fluxes were calculated from the python script written for ~\cite{Dafni:2018tvj}. The iron flux is calculated from the equations in ~\cite{DiLuzio:2021qct}.}
\label{fig:solarfluxes}
\end{figure}

\section{Discussion}
\label{sec:conclusion}
In this work, we examined the possibility of extracting information about the underlying model of the QCD axion, assuming a discovery is made in the future. If we consider only the current and upcoming landscape of axion experiments, it may be possible to learn a lot from a small number of experiments. For example, for axions in the low and intermediate mass regions with $m_a\lesssim 10^{-9} -10^{-4}$ eV, the four IR parameters ($m_a$, $g_d$, $\gagg$, $\rho_{DM}$) can be obtained from just a combination of a photon-coupling and EDM haloscopes. The electron and nucleon couplings can then potentially be probed if the axion mass is around $10^{-6}$ eV. For a heavy axion whose mass lies in the range $10^{-3} \text{ eV} \lesssim m_a \lesssim 10^1 \text{ eV}$, helioscopes have the potential to measure up to four IR parameters ($\gagg$, $g_{ae}$, $g_{aN}$, $m_a$), without any dependence on the dark matter density. These IR parameters can then be processed into element of $\datuv$ and thereby shed light on the underlying model giving rise to the observed axion.

However, it may be too ambitious to assume that all of $\datuv$ will be determined in the near future. A more modest goal is the ability to rule out some models of the QCD axion. Towards this goal, we discussed the extent to which we will be able to discriminate between KSVZ from DFSZ axion models. In the low mass range, where there are no good prospects for measuring the axion-fermion couplings $g_{af}$, the only information we expect on the underlying model is from $c^0_{a\gamma\gamma}$. This coefficient depends on the ratio of anomaly coefficients as in ~\ref{eq:directaxionphotonc}, but a range of values are possible~\cite{DiLuzio:2017pfr}. Thus experiments measuring other couplings, such as $g_{af}$, are solely need to be developed in this mass range. For axions in the intermediate mass range, DFSZ-type axions should be visible to electron haloscope experiments. This provides a means to confirm or exclude DFSZ models. In the high mass case IAXO has the potential to differentiate between DFSZ and KSVZ axions through its sensitivity to $g_{ae}$. If IAXO detects a signal from the solar axion-electron flux, the axion must be DFSZ type. If this is not seen, an upper bound can be placed on $g_{ae}$. Combining this with a measurement of the axion-nucleon coupling from the iron-57 process, DFSZ models can be ruled out. We note here that the quickest way to determine multiple UV parameters from the data occurs when we are able to identify the axion as a vanilla KSVZ or DFSZ axion. 

In this paper, we have largely considered the most optimistic scenarios to demonstrate how far the axion experimental program can go. However, it is possible that even with a positive axion signal, we can end up in a situation where we are able to learn very little of the underlying UV model. In particular, for the low and intermediate mass regimes, the experiments reaching the QCD line are all haloscopes. If the local value of $\rho_{DM}$ is smaller than expected, or if axions do not make up a significant portion of dark matter, a signal in a haloscope could be too small to be seen. 

In the previous section, we discussed situations where IAXO can fail to see an axion, leaving only photon haloscopes as discovery apparatuses. 
It is possible that even more unfortunate situations could arise in the search for the axion.
In the intermediate mass range, above the projections for the polarization haloscope ($m_a\sim10^{-4}-10^-3$ eV), the best projections are for an electron haloscope. The projections for photon haloscopes at these masses do not cover the entire range of allowed $\gagg$.  If the axion is KSVZ type and $\gagg$ is on the lower side of permitted values, it is possible to miss the axion entirely.

The pessimistic possibilities described just above identify clear gaps in the current experimental program that should be filled by new proposals in order to maximally benefit from the potential discovery of the QCD axion. On the other hand, the optimistic scenarios we have considered indicate that rapid progress in understanding the QCD axion could be possible if or when it is detected.

%%%%%%%%%%%%%%%%%%%%%%%%%%%%%%%%%%%%%%%%%%%%%%%%%%%%%%%%%%%%%%%%%%%%%%%%%%%%%%%%%%%%%%%%%%%%%%%%%%%%%%%%%%%%%%%%%%%%%%%%%%%%%%%

\begin{acknowledgments}
    We thank Nicholas Rodd for collaboration in the early stages of this work. We also thank So Chigusa and Christopher Dessert for useful discussions. The work of JML was co-funded by the European Union and supported by the Czech Ministry of Education, Youth and Sports (Project No. FORTE – CZ.02.01.01/00/22\_008/0004632). The work of HM was supported by the US DOE Contract No. DE-AC02-05CH11231, by the NSF grant PHY-2210390, by the JSPS Grant-in-Aid for Scientific Research JP23K03382, MEXT Grant-in-Aid for Transformative Research Areas (A) JP20H05850, JP20A203, BSF-2018140, by the World Premier International Research Center Initiative, MEXT, Japan, Hamamatsu Photonics, K.K, and Tokyo Dome Corporation.
\end{acknowledgments}

\appendix

%%%%%%%%%%%%%%%%%%%%%%%%%%%%%%%%%%%%%%%%%%%%%%%%%%%%%%%%%%%%%%%%%%%%%%%%%%%%%%%%%%%%%%%%%%%%%%%%%%%%%%%%%%%%%%%%%%%%%%%%%%%%%%%
\section{Axion Couplings to Mesons \& Baryons}
\label{app:couplings}
The coupling constants in~\cref{eq:axionSMcouplings} controlling the axion-pion and axion-nucleon couplings can be expressed in terms of the quark couplings $\{c^0_{a\Psi}\}$ as~\cite{DiLuzio:2020wdo}:
\begin{align}
    c_{a\pi} &= -\frac{1}{3}\bigg(c_u^0 - c_d^0 - \frac{m_d-m_u}{m_u+m_d}\bigg)\\
    %c_{a\gamma} &= g_{a\gamma}^0 - \frac{\alpha_{EM}}{2\pi f_a}\bigg(\frac{2}{3}\frac{4m_d+m_u}{m_u+m_d}\bigg)\\
    c_{ap} &= - \bigg(\frac{m_d}{m_u+m_d}\Delta u + \frac{m_u}{m_u+m_d}\Delta d\bigg) +c_u^0\Delta u + c_d^0\Delta d\\
    c_{an} &= - \bigg(\frac{m_u}{m_u+m_d}\Delta u + \frac{m_d}{m_u+m_d}\Delta d\bigg)+c_u^0\Delta u + c_d^0\Delta d\\
    c_{ae} &= c_e^0 + \frac{3\alpha_{EM}^2}{4\pi^2}\bigg(\frac{E}{N}\ln\bigg(\frac{f_a}{\mu_{IR}}\bigg)-\frac{2}{3}\frac{4m_d+m_u}{m_u+m_d}\ln\bigg(\frac{\Lambda_\chi}{\mu_{IR}}\bigg)\bigg)
\end{align}

%\begin{align}
 %   \gagg^0 = \frac{\alpha_{EM}}{2\pi f_a}\frac{E}{N}
%\end{align}
%%%%%%%%%%%%%%%%%%%%%%%%%%%%%%%%%%%%%%%%%%%%%%%%%%%%%%%%%%%%%%%%%%%%%%%%%%%%%%%%%%%%%%%%%%%%%%%%%%%%%%%%%%%%%%%%%%%%%%%%%%%%%%%

%%%%%%%%%%%%%%%%%%%%%%%%%%%%%%%%%%%%%%%%%%%%%%%%%%%%%%%%%%%%%%%%%%%%%%%%%%%%%%%%%%%%%%%%%%%%%%%%%%%%%%%%%%%%%%%%%%%%%%%%%%%%%%%
\section{Asimov analysis for IAXO}
\label{app:Asimov}
In this appendix, we estimate the uncertainty in IAXO measurements of $\gagg$ as displayed in~\cref{fig:DM_uncertainty} and discussed in~\cref{sec:axpro}. We calculate this uncertainty utilizing the Asimov approximation and using parts of the python script written for~\cite{Dafni:2018tvj}. The Asimov approximation sets the measured data set equal to the mean~\cite{Cowan:2010js}.
The likelihood function for IAXO is given by a product of Poisson distributions over the energy bins,
\begin{equation}
    \mathcal{L}(m_a,g)=\prod\limits_{i=1}^{N_{bins}}\frac{(N^i_{exp})^{N^i_{obs}}e^{-N^i_{exp}}}{N^i_{obs}!}.
\end{equation}
The dependence of the coupling in the counts is 
\begin{align}
    N^i_{exp}&=s\mathcal{N}^i(m_a) \\
    N^i_{obs}&=s\mathcal{N}^i(m_a)+N_b.
\end{align}
$s=g^4$ and $N_b$ is the (assumed flat) background.
\\We define the following test statistic to compare the case of an axion signal to the case of background only,
\begin{align}
    \tilde{\Theta}(s)&= 2\ln{\frac{\mathcal{L}(m_a,g)}{\mathcal{L}(m_a,0)}}\\
    &=2\sum_i\left[-s\mathcal{N}^i+N_{obs}^i\ln\left(1+\frac{s\mathcal{N}^i}{N_b}\right)\right]],
\end{align}
The uncertainty on the true value of $s_t=g_t^4$ is related to the curvature of $\tilde{\Theta}(s)$, and is given by
\begin{align}
    \sigma_s^{-2}&=-\frac{1}{2}\partial_s^2\tilde{\Theta}(s)|_{s=s_t}\\
    &=\sum_i\left[\frac{(\mathcal{N}^i)^2}{s_t\mathcal{N}^i+N_b}\right]
\end{align}
Using the functions defined in the code of~\cite{Dafni:2018tvj}, we then were able to calculate the uncertainty on various values for $g$ given an axion mass. We used the functions "BinnedPhotonNumberTable" and "InterpExpectedEvents" to simulate the Primakoff flux and corresponding bin counts. These functions required an input of an axion mass and coupling. We estimate the flat background as $N_b\approx 0.2$ based on the IAXO parameters given in ~\cite{IAXO:2019mpb}. Note that the uncertainty in $g$ depends on the axion mass.
%%%%%%%%%%%%%%%%%%%%%%%%%%%%%%%%%%%%%%%%%%%%%%%%%%%%%%%%%%%%%%%%%%%%%%%%%%%%%%%%%%%%%%%%%%%%%%%%%%%%%%%%%%%%%%%%%%%%%%%%%%%%%%%

\bibliographystyle{JHEP}
\bibliography{Axion_Refs}

\providecommand{\href}[2]{#2}\begingroup\raggedright\begin{thebibliography}{100}

\bibitem{Weinberg:1977ma}
S.~Weinberg, {\it {A New Light Boson?}},  {\em Phys. Rev. Lett.} {\bf 40}
  (1978) 223--226.

\bibitem{Wilczek:1977pj}
F.~Wilczek, {\it {Problem of Strong $P$ and $T$ Invariance in the Presence of
  Instantons}},  {\em Phys. Rev. Lett.} {\bf 40} (1978) 279--282.

\bibitem{Peccei:1977ur}
R.~D. Peccei and H.~R. Quinn, {\it {Constraints Imposed by CP Conservation in
  the Presence of Instantons}},  {\em Phys. Rev. D} {\bf 16} (1977) 1791--1797.

\bibitem{Peccei:1977hh}
R.~D. Peccei and H.~R. Quinn, {\it {CP Conservation in the Presence of
  Instantons}},  {\em Phys. Rev. Lett.} {\bf 38} (1977) 1440--1443.

\bibitem{Preskill:1982cy}
J.~Preskill, M.~B. Wise, and F.~Wilczek, {\it {Cosmology of the Invisible
  Axion}},  {\em Phys. Lett. B} {\bf 120} (1983) 127--132.

\bibitem{Abbott:1982af}
L.~F. Abbott and P.~Sikivie, {\it {A Cosmological Bound on the Invisible
  Axion}},  {\em Phys. Lett. B} {\bf 120} (1983) 133--136.

\bibitem{Dine:1982ah}
M.~Dine and W.~Fischler, {\it {The Not So Harmless Axion}},  {\em Phys. Lett.
  B} {\bf 120} (1983) 137--141.

\bibitem{DiLuzio:2020wdo}
L.~Di~Luzio, M.~Giannotti, E.~Nardi, and L.~Visinelli, {\it {The landscape of
  QCD axion models}},  {\em Phys. Rept.} {\bf 870} (2020) 1--117,
  [\href{http://arxiv.org/abs/2003.01100}{{\tt arXiv:2003.01100}}].

\bibitem{diCortona:2015ldu}
G.~Grilli~di Cortona, E.~Hardy, J.~Pardo~Vega, and G.~Villadoro, {\it {The QCD
  axion, precisely}},  {\em JHEP} {\bf 01} (2016) 034,
  [\href{http://arxiv.org/abs/1511.02867}{{\tt arXiv:1511.02867}}].

\bibitem{Svrcek:2006yi}
P.~Svrcek and E.~Witten, {\it {Axions In String Theory}},  {\em JHEP} {\bf 06}
  (2006) 051, [\href{http://arxiv.org/abs/hep-th/0605206}{{\tt
  hep-th/0605206}}].

\bibitem{Zhitnitsky:1980tq}
A.~R. Zhitnitsky, {\it {On Possible Suppression of the Axion Hadron
  Interactions. (In Russian)}},  {\em Sov. J. Nucl. Phys.} {\bf 31} (1980) 260.

\bibitem{Dine:1981rt}
M.~Dine, W.~Fischler, and M.~Srednicki, {\it {A Simple Solution to the Strong
  CP Problem with a Harmless Axion}},  {\em Phys. Lett. B} {\bf 104} (1981)
  199--202.

\bibitem{Kim:1979if}
J.~E. Kim, {\it {Weak Interaction Singlet and Strong CP Invariance}},  {\em
  Phys. Rev. Lett.} {\bf 43} (1979) 103.

\bibitem{Shifman:1979if}
M.~A. Shifman, A.~I. Vainshtein, and V.~I. Zakharov, {\it {Can Confinement
  Ensure Natural CP Invariance of Strong Interactions?}},  {\em Nucl. Phys. B}
  {\bf 166} (1980) 493--506.

\bibitem{DiLuzio:2016sur}
L.~Di~Luzio, J.~F. Kamenik, and M.~Nardecchia, {\it {Implications of
  perturbative unitarity for scalar di-boson resonance searches at LHC}},  {\em
  Eur. Phys. J. C} {\bf 77} (2017), no.~1 30,
  [\href{http://arxiv.org/abs/1604.05746}{{\tt arXiv:1604.05746}}].

\bibitem{DiLuzio:2017chi}
L.~Di~Luzio and M.~Nardecchia, {\it {What is the scale of new physics behind
  the $B$-flavour anomalies?}},  {\em Eur. Phys. J. C} {\bf 77} (2017), no.~8
  536, [\href{http://arxiv.org/abs/1706.01868}{{\tt arXiv:1706.01868}}].

\bibitem{Bjorkeroth:2019jtx}
F.~Bj\"orkeroth, L.~Di~Luzio, F.~Mescia, E.~Nardi, P.~Panci, and R.~Ziegler,
  {\it {Axion-electron decoupling in nucleophobic axion models}},  {\em Phys.
  Rev. D} {\bf 101} (2020), no.~3 035027,
  [\href{http://arxiv.org/abs/1907.06575}{{\tt arXiv:1907.06575}}].

\bibitem{Dror:2020zru}
J.~A. Dror and J.~M. Leedom, {\it {Cosmological Tension of Ultralight Axion
  Dark Matter and its Solutions}},  {\em Phys. Rev. D} {\bf 102} (2020), no.~11
  115030, [\href{http://arxiv.org/abs/2008.02279}{{\tt arXiv:2008.02279}}].

\bibitem{Agrawal:2017cmd}
P.~Agrawal, J.~Fan, M.~Reece, and L.-T. Wang, {\it {Experimental Targets for
  Photon Couplings of the QCD Axion}},  {\em JHEP} {\bf 02} (2018) 006,
  [\href{http://arxiv.org/abs/1709.06085}{{\tt arXiv:1709.06085}}].

\bibitem{Agrawal:2018mkd}
P.~Agrawal, J.~Fan, and M.~Reece, {\it {Clockwork Axions in Cosmology: Is
  Chromonatural Inflation Chrononatural?}},  {\em JHEP} {\bf 10} (2018) 193,
  [\href{http://arxiv.org/abs/1806.09621}{{\tt arXiv:1806.09621}}].

\bibitem{Choi:2015fiu}
K.~Choi and S.~H. Im, {\it {Realizing the relaxion from multiple axions and its
  UV completion with high scale supersymmetry}},  {\em JHEP} {\bf 01} (2016)
  149, [\href{http://arxiv.org/abs/1511.00132}{{\tt arXiv:1511.00132}}].

\bibitem{Farina:2016tgd}
M.~Farina, D.~Pappadopulo, F.~Rompineve, and A.~Tesi, {\it {The photo-philic
  QCD axion}},  {\em JHEP} {\bf 01} (2017) 095,
  [\href{http://arxiv.org/abs/1611.09855}{{\tt arXiv:1611.09855}}].

\bibitem{Coy:2017yex}
R.~Coy, M.~Frigerio, and M.~Ibe, {\it {Dynamical Clockwork Axions}},  {\em
  JHEP} {\bf 10} (2017) 002, [\href{http://arxiv.org/abs/1706.04529}{{\tt
  arXiv:1706.04529}}].

\bibitem{Marques-Tavares:2018cwm}
G.~Marques-Tavares and M.~Teo, {\it {Light axions with large hadronic
  couplings}},  {\em JHEP} {\bf 05} (2018) 180,
  [\href{http://arxiv.org/abs/1803.07575}{{\tt arXiv:1803.07575}}].

\bibitem{Babu:1994id}
K.~S. Babu, S.~M. Barr, and D.~Seckel, {\it {Axion dissipation through the
  mixing of Goldstone bosons}},  {\em Phys. Lett. B} {\bf 336} (1994) 213--220,
  [\href{http://arxiv.org/abs/hep-ph/9406308}{{\tt hep-ph/9406308}}].

\bibitem{Cicoli:2012sz}
M.~Cicoli, M.~Goodsell, and A.~Ringwald, {\it {The type IIB string axiverse and
  its low-energy phenomenology}},  {\em JHEP} {\bf 10} (2012) 146,
  [\href{http://arxiv.org/abs/1206.0819}{{\tt arXiv:1206.0819}}].

\bibitem{Higaki:2014qua}
T.~Higaki, N.~Kitajima, and F.~Takahashi, {\it {Hidden axion dark matter
  decaying through mixing with QCD axion and the 3.5 keV X-ray line}},  {\em
  JCAP} {\bf 12} (2014) 004, [\href{http://arxiv.org/abs/1408.3936}{{\tt
  arXiv:1408.3936}}].

\bibitem{Bachlechner:2014hsa}
T.~C. Bachlechner, M.~Dias, J.~Frazer, and L.~McAllister, {\it {Chaotic
  inflation with kinetic alignment of axion fields}},  {\em Phys. Rev. D} {\bf
  91} (2015), no.~2 023520, [\href{http://arxiv.org/abs/1404.7496}{{\tt
  arXiv:1404.7496}}].

\bibitem{Shiu:2015uva}
G.~Shiu, W.~Staessens, and F.~Ye, {\it {Widening the Axion Window via Kinetic
  and St\"uckelberg Mixings}},  {\em Phys. Rev. Lett.} {\bf 115} (2015) 181601,
  [\href{http://arxiv.org/abs/1503.01015}{{\tt arXiv:1503.01015}}].

\bibitem{Shiu:2015xda}
G.~Shiu, W.~Staessens, and F.~Ye, {\it {Large Field Inflation from Axion
  Mixing}},  {\em JHEP} {\bf 06} (2015) 026,
  [\href{http://arxiv.org/abs/1503.02965}{{\tt arXiv:1503.02965}}].

\bibitem{Agrawal:2017eqm}
P.~Agrawal, G.~Marques-Tavares, and W.~Xue, {\it {Opening up the QCD axion
  window}},  {\em JHEP} {\bf 03} (2018) 049,
  [\href{http://arxiv.org/abs/1708.05008}{{\tt arXiv:1708.05008}}].

\bibitem{Daido:2018dmu}
R.~Daido, F.~Takahashi, and N.~Yokozaki, {\it {Enhanced
  axion\textendash{}photon coupling in GUT with hidden photon}},  {\em Phys.
  Lett. B} {\bf 780} (2018) 538--542,
  [\href{http://arxiv.org/abs/1801.10344}{{\tt arXiv:1801.10344}}].

\bibitem{Hook:2018jle}
A.~Hook, {\it {Solving the Hierarchy Problem Discretely}},  {\em Phys. Rev.
  Lett.} {\bf 120} (2018), no.~26 261802,
  [\href{http://arxiv.org/abs/1802.10093}{{\tt arXiv:1802.10093}}].

\bibitem{DiLuzio:2021pxd}
L.~Di~Luzio, B.~Gavela, P.~Quilez, and A.~Ringwald, {\it {An even lighter QCD
  axion}},  {\em JHEP} {\bf 05} (2021) 184,
  [\href{http://arxiv.org/abs/2102.00012}{{\tt arXiv:2102.00012}}].

\bibitem{DiLuzio:2021gos}
L.~Di~Luzio, B.~Gavela, P.~Quilez, and A.~Ringwald, {\it {Dark matter from an
  even lighter QCD axion: trapped misalignment}},  {\em JCAP} {\bf 10} (2021)
  001, [\href{http://arxiv.org/abs/2102.01082}{{\tt arXiv:2102.01082}}].

\bibitem{Hebecker:2015tzo}
A.~Hebecker, J.~Moritz, A.~Westphal, and L.~T. Witkowski, {\it {Axion Monodromy
  Inflation with Warped KK-Modes}},  {\em Phys. Lett. B} {\bf 754} (2016)
  328--334, [\href{http://arxiv.org/abs/1512.04463}{{\tt arXiv:1512.04463}}].
  [Erratum: Phys.Lett.B 767, 493--493 (2017)].

\bibitem{Hebecker:2018yxs}
A.~Hebecker, S.~Leonhardt, J.~Moritz, and A.~Westphal, {\it {Thraxions:
  Ultralight Throat Axions}},  {\em JHEP} {\bf 04} (2019) 158,
  [\href{http://arxiv.org/abs/1812.03999}{{\tt arXiv:1812.03999}}].

\bibitem{Carta:2021uwv}
F.~Carta, A.~Mininno, N.~Righi, and A.~Westphal, {\it {Thraxions: towards full
  string models}},  {\em JHEP} {\bf 01} (2022) 082,
  [\href{http://arxiv.org/abs/2110.02963}{{\tt arXiv:2110.02963}}].

\bibitem{Arvanitaki:2009fg}
A.~Arvanitaki, S.~Dimopoulos, S.~Dubovsky, N.~Kaloper, and J.~March-Russell,
  {\it {String Axiverse}},  {\em Phys. Rev. D} {\bf 81} (2010) 123530,
  [\href{http://arxiv.org/abs/0905.4720}{{\tt arXiv:0905.4720}}].

\bibitem{Gaillard:2005gj}
M.~K. Gaillard and B.~Kain, {\it {Is the universal string axion the QCD
  axion?}},  {\em Nucl. Phys. B} {\bf 734} (2006) 116--137,
  [\href{http://arxiv.org/abs/hep-th/0510190}{{\tt hep-th/0510190}}].

\bibitem{Fox:2004kb}
P.~Fox, A.~Pierce, and S.~D. Thomas, {\it {Probing a QCD string axion with
  precision cosmological measurements}},
  \href{http://arxiv.org/abs/hep-th/0409059}{{\tt hep-th/0409059}}.

\bibitem{Lee:2022mnc}
Y.~Lee, B.~Yang, H.~Yoon, M.~Ahn, H.~Park, B.~Min, D.~Kim, and J.~Yoo, {\it
  {Searching for Invisible Axion Dark Matter with an 18~T Magnet Haloscope}},
  {\em Phys. Rev. Lett.} {\bf 128} (2022), no.~24 241805,
  [\href{http://arxiv.org/abs/2206.08845}{{\tt arXiv:2206.08845}}].

\bibitem{Chigusa:2020gfs}
S.~Chigusa, T.~Moroi, and K.~Nakayama, {\it {Detecting light boson dark matter
  through conversion into a magnon}},  {\em Phys. Rev. D} {\bf 101} (2020),
  no.~9 096013, [\href{http://arxiv.org/abs/2001.10666}{{\tt
  arXiv:2001.10666}}].

\bibitem{Mitridate:2020kly}
A.~Mitridate, T.~Trickle, Z.~Zhang, and K.~M. Zurek, {\it {Detectability of
  Axion Dark Matter with Phonon Polaritons and Magnons}},  {\em Phys. Rev. D}
  {\bf 102} (2020), no.~9 095005, [\href{http://arxiv.org/abs/2005.10256}{{\tt
  arXiv:2005.10256}}].

\bibitem{Sikivie:2014lha}
P.~Sikivie, {\it {Axion Dark Matter Detection using Atomic Transitions}},  {\em
  Phys. Rev. Lett.} {\bf 113} (2014), no.~20 201301,
  [\href{http://arxiv.org/abs/1409.2806}{{\tt arXiv:1409.2806}}]. [Erratum:
  Phys.Rev.Lett. 125, 029901 (2020)].

\bibitem{Berlin:2023ubt}
A.~Berlin, A.~J. Millar, T.~Trickle, and K.~Zhou, {\it {Physical Signatures of
  Fermion-Coupled Axion Dark Matter}},
  \href{http://arxiv.org/abs/2312.11601}{{\tt arXiv:2312.11601}}.

\bibitem{Graham:2013gfa}
P.~W. Graham and S.~Rajendran, {\it {New Observables for Direct Detection of
  Axion Dark Matter}},  {\em Phys. Rev. D} {\bf 88} (2013) 035023,
  [\href{http://arxiv.org/abs/1306.6088}{{\tt arXiv:1306.6088}}].

\bibitem{Budker:2013hfa}
D.~Budker, P.~W. Graham, M.~Ledbetter, S.~Rajendran, and A.~Sushkov, {\it
  {Proposal for a Cosmic Axion Spin Precession Experiment (CASPEr)}},  {\em
  Phys. Rev. X} {\bf 4} (2014), no.~2 021030,
  [\href{http://arxiv.org/abs/1306.6089}{{\tt arXiv:1306.6089}}].

\bibitem{JacksonKimball:2017elr}
D.~F. Jackson~Kimball et~al., {\it {Overview of the Cosmic Axion Spin
  Precession Experiment (CASPEr)}},  {\em Springer Proc. Phys.} {\bf 245}
  (2020) 105--121, [\href{http://arxiv.org/abs/1711.08999}{{\tt
  arXiv:1711.08999}}].

\bibitem{Stadnik:2014xja}
Y.~V. Stadnik and V.~V. Flambaum, {\it {Nuclear spin-dependent interactions:
  Searches for WIMP, Axion and Topological Defect Dark Matter, and Tests of
  Fundamental Symmetries}},  {\em Eur. Phys. J. C} {\bf 75} (2015), no.~3 110,
  [\href{http://arxiv.org/abs/1408.2184}{{\tt arXiv:1408.2184}}].

\bibitem{Chigusa:2023szl}
S.~Chigusa, D.~Kondo, H.~Murayama, R.~Okabe, and H.~Sudo, {\it {Axion detection
  via superfluid $^3$He ferromagnetic phase and quantum measurement
  techniques}},  \href{http://arxiv.org/abs/2309.09160}{{\tt
  arXiv:2309.09160}}.

\bibitem{Chigusa:2023hmz}
S.~Chigusa, T.~Moroi, K.~Nakayama, and T.~Sichanugrist, {\it {Dark matter
  detection using nuclear magnetization in magnet with hyperfine interaction}},
   {\em Phys. Rev. D} {\bf 108} (2023), no.~9 095007,
  [\href{http://arxiv.org/abs/2307.08577}{{\tt arXiv:2307.08577}}].

\bibitem{Garcon_2017}
A.~Garcon, D.~Aybas, J.~W. Blanchard, G.~Centers, N.~L. Figueroa, P.~W. Graham,
  D.~F.~J. Kimball, S.~Rajendran, M.~G. Sendra, A.~O. Sushkov, L.~Trahms,
  T.~Wang, A.~Wickenbrock, T.~Wu, and D.~Budker, {\it The cosmic axion spin
  precession experiment (casper): a dark-matter search with nuclear magnetic
  resonance},  {\em Quantum Science and Technology} {\bf 3} (Dec., 2017)
  014008.

\bibitem{PhysRevD.42.1847}
J.~Hong, J.~E. Kim, and P.~Sikivie, {\it Nuclear dipole radiation from
  $\overline{\ensuremath{\theta}}$ oscillations},  {\em Phys. Rev. D} {\bf 42}
  (Sep, 1990) 1847--1850.

\bibitem{Berlin:2022mia}
A.~Berlin and K.~Zhou, {\it {Discovering QCD-Coupled Axion Dark Matter with
  Polarization Haloscopes}},  \href{http://arxiv.org/abs/2209.12901}{{\tt
  arXiv:2209.12901}}.

\bibitem{CAST:2008ixs}
{\bf CAST} Collaboration, E.~Arik et~al., {\it {Probing eV-scale axions with
  CAST}},  {\em JCAP} {\bf 02} (2009) 008,
  [\href{http://arxiv.org/abs/0810.4482}{{\tt arXiv:0810.4482}}].

\bibitem{Redondo:2013wwa}
J.~Redondo, {\it {Solar axion flux from the axion-electron coupling}},  {\em
  JCAP} {\bf 12} (2013) 008, [\href{http://arxiv.org/abs/1310.0823}{{\tt
  arXiv:1310.0823}}].

\bibitem{Jaeckel:2018mbn}
J.~Jaeckel and L.~J. Thormaehlen, {\it {Distinguishing Axion Models with
  IAXO}},  {\em JCAP} {\bf 03} (2019) 039,
  [\href{http://arxiv.org/abs/1811.09278}{{\tt arXiv:1811.09278}}].

\bibitem{DiLuzio:2021qct}
L.~Di~Luzio et~al., {\it {Probing the axion\textendash{}nucleon coupling with
  the next generation of~axion helioscopes}},  {\em Eur. Phys. J. C} {\bf 82}
  (2022), no.~2 120, [\href{http://arxiv.org/abs/2111.06407}{{\tt
  arXiv:2111.06407}}].

\bibitem{Dafni:2018tvj}
T.~Dafni, C.~A.~J. O'Hare, B.~Laki\'c, J.~Gal\'an, F.~J. Iguaz, I.~G.
  Irastorza, K.~Jakov\v{c}ic, G.~Luz\'on, J.~Redondo, and E.~Ruiz~Ch\'oliz,
  {\it {Weighing the solar axion}},  {\em Phys. Rev. D} {\bf 99} (2019), no.~3
  035037, [\href{http://arxiv.org/abs/1811.09290}{{\tt arXiv:1811.09290}}].

\bibitem{Hoof:2021mld}
S.~Hoof, J.~Jaeckel, and L.~J. Thormaehlen, {\it {Quantifying uncertainties in
  the solar axion flux and their impact on determining axion model
  parameters}},  \href{http://arxiv.org/abs/2101.08789}{{\tt
  arXiv:2101.08789}}.

\bibitem{OHare:2020wum}
C.~A.~J. O'Hare, A.~Caputo, A.~J. Millar, and E.~Vitagliano, {\it {Axion
  helioscopes as solar magnetometers}},  {\em Phys. Rev. D} {\bf 102} (2020),
  no.~4 043019, [\href{http://arxiv.org/abs/2006.10415}{{\tt
  arXiv:2006.10415}}].

\bibitem{Moriyama:1995bz}
S.~Moriyama, {\it {A Proposal to search for a monochromatic component of solar
  axions using Fe-57}},  {\em Phys. Rev. Lett.} {\bf 75} (1995) 3222--3225,
  [\href{http://arxiv.org/abs/hep-ph/9504318}{{\tt hep-ph/9504318}}].

\bibitem{Moriyama:1998tx}
S.~Moriyama, {\it {Proposal to search for a monochromatic component of solar
  axions using Fe-57}},  {\em Nucl. Phys. B Proc. Suppl.} {\bf 72} (1999)
  183--186, [\href{http://arxiv.org/abs/hep-ex/9805032}{{\tt hep-ex/9805032}}].

\bibitem{Krcmar:1998xn}
M.~Krcmar, Z.~Krecak, M.~Stipcevic, A.~Ljubicic, and D.~A. Bradley, {\it
  {Search for invisible axions using Fe-57}},  {\em Phys. Lett. B} {\bf 442}
  (1998) 38, [\href{http://arxiv.org/abs/nucl-ex/9801005}{{\tt
  nucl-ex/9801005}}].

\bibitem{Read:2014qva}
J.~I. Read, {\it {The Local Dark Matter Density}},  {\em J. Phys. G} {\bf 41}
  (2014) 063101, [\href{http://arxiv.org/abs/1404.1938}{{\tt
  arXiv:1404.1938}}].

\bibitem{Green:2017odb}
A.~M. Green, {\it {Astrophysical uncertainties on the local dark matter
  distribution and direct detection experiments}},  {\em J. Phys. G} {\bf 44}
  (2017), no.~8 084001, [\href{http://arxiv.org/abs/1703.10102}{{\tt
  arXiv:1703.10102}}].

\bibitem{Dessert:2021bkv}
C.~Dessert, A.~J. Long, and B.~R. Safdi, {\it {No Evidence for Axions from
  Chandra Observation of the Magnetic White Dwarf RE J0317-853}},  {\em Phys.
  Rev. Lett.} {\bf 128} (2022), no.~7 071102,
  [\href{http://arxiv.org/abs/2104.12772}{{\tt arXiv:2104.12772}}].

\bibitem{Noordhuis:2023wid}
D.~Noordhuis, A.~Prabhu, C.~Weniger, and S.~J. Witte, {\it {Axion Clouds around
  Neutron Stars}},  \href{http://arxiv.org/abs/2307.11811}{{\tt
  arXiv:2307.11811}}.

\bibitem{Batllori:2023gwy}
J.~M. Batllori, Y.~Gu, D.~Horns, M.~Maroudas, and J.~Ulrichs, {\it {Searching
  for weakly interacting sub-eV particles with a fiber interferometer in a
  strong magnetic field}},  {\em Phys. Rev. D} {\bf 109} (2024), no.~12 123001,
  [\href{http://arxiv.org/abs/2305.12969}{{\tt arXiv:2305.12969}}].

\bibitem{Chaudhuri:2014dla}
S.~Chaudhuri, P.~W. Graham, K.~Irwin, J.~Mardon, S.~Rajendran, and Y.~Zhao,
  {\it {Radio for hidden-photon dark matter detection}},  {\em Phys. Rev. D}
  {\bf 92} (2015), no.~7 075012, [\href{http://arxiv.org/abs/1411.7382}{{\tt
  arXiv:1411.7382}}].

\bibitem{Berlin:2020vrk}
A.~Berlin, R.~T. D'Agnolo, S.~A.~R. Ellis, and K.~Zhou, {\it {Heterodyne
  broadband detection of axion dark matter}},  {\em Phys. Rev. D} {\bf 104}
  (2021), no.~11 L111701, [\href{http://arxiv.org/abs/2007.15656}{{\tt
  arXiv:2007.15656}}].

\bibitem{Zhang:2021bpa}
Z.~Zhang, D.~Horns, and O.~Ghosh, {\it {Search for dark matter with an LC
  circuit}},  {\em Phys. Rev. D} {\bf 106} (2022), no.~2 023003,
  [\href{http://arxiv.org/abs/2111.04541}{{\tt arXiv:2111.04541}}].

\bibitem{Nagano:2019rbw}
K.~Nagano, T.~Fujita, Y.~Michimura, and I.~Obata, {\it {Axion Dark Matter
  Search with Interferometric Gravitational Wave Detectors}},  {\em Phys. Rev.
  Lett.} {\bf 123} (2019), no.~11 111301,
  [\href{http://arxiv.org/abs/1903.02017}{{\tt arXiv:1903.02017}}].

\bibitem{Michimura:2019qxr}
Y.~Michimura, Y.~Oshima, T.~Watanabe, T.~Kawasaki, H.~Takeda, M.~Ando,
  K.~Nagano, I.~Obata, and T.~Fujita, {\it {DANCE: Dark matter Axion search
  with riNg Cavity Experiment}},  {\em J. Phys. Conf. Ser.} {\bf 1468} (2020),
  no.~1 012032, [\href{http://arxiv.org/abs/1911.05196}{{\tt
  arXiv:1911.05196}}].

\bibitem{Liu:2018icu}
H.~Liu, B.~D. Elwood, M.~Evans, and J.~Thaler, {\it {Searching for Axion Dark
  Matter with Birefringent Cavities}},  {\em Phys. Rev. D} {\bf 100} (2019),
  no.~2 023548, [\href{http://arxiv.org/abs/1809.01656}{{\tt
  arXiv:1809.01656}}].

\bibitem{Dror:2022xpi}
J.~A. Dror, S.~Gori, J.~M. Leedom, and N.~L. Rodd, {\it {On the Sensitivity of
  Spin-Precession Axion Experiments}},
  \href{http://arxiv.org/abs/2210.06481}{{\tt arXiv:2210.06481}}.

\bibitem{DiLuzio:2017pfr}
L.~Di~Luzio, F.~Mescia, and E.~Nardi, {\it {Window for preferred axion
  models}},  {\em Phys. Rev. D} {\bf 96} (2017), no.~7 075003,
  [\href{http://arxiv.org/abs/1705.05370}{{\tt arXiv:1705.05370}}].

\bibitem{Asztalos_2010}
S.~J. Asztalos, G.~Carosi, C.~Hagmann, D.~Kinion, K.~van Bibber, M.~Hotz, L.~J.
  Rosenberg, G.~Rybka, J.~Hoskins, J.~Hwang, and et~al., {\it Squid-based
  microwave cavity search for dark-matter axions},  {\em Physical Review
  Letters} {\bf 104} (Jan, 2010).

\bibitem{Du:2018uak}
{\bf ADMX} Collaboration, N.~Du et~al., {\it {A Search for Invisible Axion Dark
  Matter with the Axion Dark Matter Experiment}},  {\em Phys. Rev. Lett.} {\bf
  120} (2018), no.~15 151301, [\href{http://arxiv.org/abs/1804.05750}{{\tt
  arXiv:1804.05750}}].

\bibitem{Braine:2019fqb}
{\bf ADMX} Collaboration, T.~Braine et~al., {\it {Extended Search for the
  Invisible Axion with the Axion Dark Matter Experiment}},  {\em Phys. Rev.
  Lett.} {\bf 124} (2020), no.~10 101303,
  [\href{http://arxiv.org/abs/1910.08638}{{\tt arXiv:1910.08638}}].

\bibitem{Boutan:2018uoc}
{\bf ADMX} Collaboration, C.~Boutan et~al., {\it {Piezoelectrically Tuned
  Multimode Cavity Search for Axion Dark Matter}},  {\em Phys. Rev. Lett.} {\bf
  121} (2018), no.~26 261302, [\href{http://arxiv.org/abs/1901.00920}{{\tt
  arXiv:1901.00920}}].

\bibitem{Stern:2016bbw}
I.~Stern, {\it {ADMX Status}},  {\em PoS} {\bf ICHEP2016} (2016) 198,
  [\href{http://arxiv.org/abs/1612.08296}{{\tt arXiv:1612.08296}}].

\bibitem{Crisosto:2019fcj}
N.~Crisosto, P.~Sikivie, N.~S. Sullivan, D.~B. Tanner, J.~Yang, and G.~Rybka,
  {\it {ADMX SLIC: Results from a Superconducting $LC$ Circuit Investigating
  Cold Axions}},  {\em Phys. Rev. Lett.} {\bf 124} (2020), no.~24 241101,
  [\href{http://arxiv.org/abs/1911.05772}{{\tt arXiv:1911.05772}}].

\bibitem{ADMXwhite:2020}
A.~Sonnenschein, {\it {Snowmass2021 - Letter of Interest Axion Dark Matter
  eXperiment (ADMX) 2-4 GHz}},  {\em .}

\bibitem{Lawson:2019brd}
M.~Lawson, A.~J. Millar, M.~Pancaldi, E.~Vitagliano, and F.~Wilczek, {\it
  {Tunable axion plasma haloscopes}},  {\em Phys. Rev. Lett.} {\bf 123} (2019),
  no.~14 141802, [\href{http://arxiv.org/abs/1904.11872}{{\tt
  arXiv:1904.11872}}].

\bibitem{FLASH2022}
C.~Gatti, ``Flash, a proposal for a 100-300 mhz haloscope.'' Physics
  Opportunities at 100-500 MHz Haloscopes, 2022.

\bibitem{Alesini:2019ajt}
D.~Alesini et~al., {\it {Galactic axions search with a superconducting resonant
  cavity}},  {\em Phys. Rev. D} {\bf 99} (2019), no.~10 101101,
  [\href{http://arxiv.org/abs/1903.06547}{{\tt arXiv:1903.06547}}].

\bibitem{Alesini:2020vny}
D.~Alesini et~al., {\it {Search for invisible axion dark matter of mass
  m$_a=43~\mu$eV with the QUAX--$a\gamma$ experiment}},  {\em Phys. Rev. D}
  {\bf 103} (2021), no.~10 102004, [\href{http://arxiv.org/abs/2012.09498}{{\tt
  arXiv:2012.09498}}].

\bibitem{Alesini:2022lnp}
D.~Alesini et~al., {\it {Search for Galactic axions with a high-Q dielectric
  cavity}},  {\em Phys. Rev. D} {\bf 106} (2022), no.~5 052007,
  [\href{http://arxiv.org/abs/2208.12670}{{\tt arXiv:2208.12670}}].

\bibitem{QUAX:2023gop}
{\bf QUAX} Collaboration, R.~Di~Vora et~al., {\it {Search for galactic axions
  with a traveling wave parametric amplifier}},  {\em Phys. Rev. D} {\bf 108}
  (2023), no.~6 062005, [\href{http://arxiv.org/abs/2304.07505}{{\tt
  arXiv:2304.07505}}].

\bibitem{QUAX:2024fut}
{\bf QUAX} Collaboration, A.~Rettaroli et~al., {\it {Search for axion dark
  matter with the QUAX\textendash{}LNF tunable haloscope}},  {\em Phys. Rev. D}
  {\bf 110} (2024), no.~2 022008, [\href{http://arxiv.org/abs/2402.19063}{{\tt
  arXiv:2402.19063}}].

\bibitem{QUAX2021}
A.~Rettaroli, ``Probing the axion-photon interaction with quax experiment:
  status and perspectives.'' 16TH Patras Workshop on Axions, WIMPs, and WISPs,
  2021.

\bibitem{Ahyoune:2023gfw}
S.~Ahyoune et~al., {\it {A Proposal for a Low-Frequency Axion Search in the
  1\textendash{}2 $\mu$eV Range and Below with the BabyIAXO Magnet}},  {\em
  Annalen Phys.} {\bf 535} (2023), no.~12 2300326,
  [\href{http://arxiv.org/abs/2306.17243}{{\tt arXiv:2306.17243}}].

\bibitem{DeMiguel:2023nmz}
{\bf DALI} Collaboration, J.~De~Miguel, J.~F. Hern\'andez-Cabrera,
  E.~Hern\'andez-Su\'arez, E.~Joven-\'Alvarez, C.~Otani, and J.~A. Rubi\~no
  Mart\'\i{}n, {\it {Discovery prospects with the Dark-photons \& Axion-like
  particles Interferometer}},  {\em Phys. Rev. D} {\bf 109} (2024), no.~6
  062002, [\href{http://arxiv.org/abs/2303.03997}{{\tt arXiv:2303.03997}}].

\bibitem{McAllister:2017lkb}
B.~T. McAllister, G.~Flower, J.~Kruger, E.~N. Ivanov, M.~Goryachev,
  J.~Bourhill, and M.~E. Tobar, {\it {The ORGAN Experiment: An axion haloscope
  above 15 GHz}},  {\em Phys. Dark Univ.} {\bf 18} (2017) 67--72,
  [\href{http://arxiv.org/abs/1706.00209}{{\tt arXiv:1706.00209}}].

\bibitem{Quiskamp:2022pks}
A.~P. Quiskamp, B.~T. McAllister, P.~Altin, E.~N. Ivanov, M.~Goryachev, and
  M.~E. Tobar, {\it {Direct search for dark matter axions excluding ALP
  cogenesis in the 63- to 67-\ensuremath{\mu}eV range with the ORGAN
  experiment}},  {\em Sci. Adv.} {\bf 8} (2022), no.~27 abq3765,
  [\href{http://arxiv.org/abs/2203.12152}{{\tt arXiv:2203.12152}}].

\bibitem{Quiskamp:2023ehr}
A.~Quiskamp, B.~T. McAllister, P.~Altin, E.~N. Ivanov, M.~Goryachev, and M.~E.
  Tobar, {\it {Exclusion of Axionlike-Particle Cogenesis Dark Matter in a Mass
  Window above 100\,\,\ensuremath{\mu}eV}},  {\em Phys. Rev. Lett.} {\bf 132}
  (2024), no.~3 031601, [\href{http://arxiv.org/abs/2310.00904}{{\tt
  arXiv:2310.00904}}].

\bibitem{Quiskamp:2024oet}
A.~P. Quiskamp, G.~Flower, S.~Samuels, B.~T. McAllister, P.~Altin, E.~N.
  Ivanov, M.~Goryachev, and M.~E. Tobar, {\it {Near-quantum limited axion dark
  matter search with the ORGAN experiment around 26 $\mu$eV}},
  \href{http://arxiv.org/abs/2407.18586}{{\tt arXiv:2407.18586}}.

\bibitem{Lee:2020cfj}
S.~Lee, S.~Ahn, J.~Choi, B.~R. Ko, and Y.~K. Semertzidis, {\it {Axion Dark
  Matter Search around 6.7 $\mu$eV}},  {\em Phys. Rev. Lett.} {\bf 124} (2020),
  no.~10 101802, [\href{http://arxiv.org/abs/2001.05102}{{\tt
  arXiv:2001.05102}}].

\bibitem{Jeong:2020cwz}
J.~Jeong, S.~Youn, S.~Bae, J.~Kim, T.~Seong, J.~E. Kim, and Y.~K. Semertzidis,
  {\it {Search for Invisible Axion Dark Matter with a Multiple-Cell
  Haloscope}},  {\em Phys. Rev. Lett.} {\bf 125} (2020), no.~22 221302,
  [\href{http://arxiv.org/abs/2008.10141}{{\tt arXiv:2008.10141}}].

\bibitem{CAPP:2020utb}
{\bf CAPP} Collaboration, O.~Kwon et~al., {\it {First Results from an Axion
  Haloscope at CAPP around 10.7 $\mu$eV}},  {\em Phys. Rev. Lett.} {\bf 126}
  (2021), no.~19 191802, [\href{http://arxiv.org/abs/2012.10764}{{\tt
  arXiv:2012.10764}}].

\bibitem{Yoon:2022gzp}
H.~Yoon, M.~Ahn, B.~Yang, Y.~Lee, D.~Kim, H.~Park, B.~Min, and J.~Yoo, {\it
  {Axion haloscope using an 18~T high temperature superconducting magnet}},
  {\em Phys. Rev. D} {\bf 106} (2022), no.~9 092007,
  [\href{http://arxiv.org/abs/2206.12271}{{\tt arXiv:2206.12271}}].

\bibitem{Kim:2022hmg}
J.~Kim et~al., {\it {Near-Quantum-Noise Axion Dark Matter Search at CAPP around
  9.5 $\mu$eV}},  {\em Phys. Rev. Lett.} {\bf 130} (2023), no.~9 091602,
  [\href{http://arxiv.org/abs/2207.13597}{{\tt arXiv:2207.13597}}].

\bibitem{Yi:2022fmn}
A.~K. Yi et~al., {\it {Axion Dark Matter Search around 4.55 $\mu$eV with
  Dine-Fischler-Srednicki-Zhitnitskii Sensitivity}},  {\em Phys. Rev. Lett.}
  {\bf 130} (2023), no.~7 071002, [\href{http://arxiv.org/abs/2210.10961}{{\tt
  arXiv:2210.10961}}].

\bibitem{Adair:2022rtw}
C.~M. Adair et~al., {\it {Search for Dark Matter Axions with CAST-CAPP}},  {\em
  Nature Commun.} {\bf 13} (2022), no.~1 6180,
  [\href{http://arxiv.org/abs/2211.02902}{{\tt arXiv:2211.02902}}].

\bibitem{Kim:2023vpo}
Y.~Kim et~al., {\it {Experimental Search for Invisible Dark Matter Axions
  around 22 $\mu$eV}},  {\em Phys. Rev. Lett.} {\bf 133} (2024), no.~5 051802,
  [\href{http://arxiv.org/abs/2312.11003}{{\tt arXiv:2312.11003}}].

\bibitem{Yang:2023yry}
B.~Yang, H.~Yoon, M.~Ahn, Y.~Lee, and J.~Yoo, {\it {Extended Axion Dark Matter
  Search Using the CAPP18T Haloscope}},  {\em Phys. Rev. Lett.} {\bf 131}
  (2023), no.~8 081801, [\href{http://arxiv.org/abs/2308.09077}{{\tt
  arXiv:2308.09077}}].

\bibitem{CAPP:2024dtx}
{\bf CAPP} Collaboration, S.~Ahn et~al., {\it {Extensive Search for Axion Dark
  Matter over 1~GHz with CAPP Main Axion Experiment}},  {\em Phys. Rev. X} {\bf
  14} (2024), no.~3 031023, [\href{http://arxiv.org/abs/2402.12892}{{\tt
  arXiv:2402.12892}}].

\bibitem{Beurthey:2020yuq}
S.~Beurthey et~al., {\it {MADMAX Status Report}},
  \href{http://arxiv.org/abs/2003.10894}{{\tt arXiv:2003.10894}}.

\bibitem{Braggio:2017oyt}
C.~Braggio et~al., {\it {Axion dark matter detection by laser induced
  fluorescence in rare-earth doped materials}},  {\em Sci. Rep.} {\bf 7}
  (2017), no.~1 15168, [\href{http://arxiv.org/abs/1707.06103}{{\tt
  arXiv:1707.06103}}].

\bibitem{Vergados:2018qdb}
J.~D. Vergados, F.~T. Avignone, S.~Cohen, and R.~J. Creswick, {\it {Axion
  Detection via Atomic Excitations}},
  \href{http://arxiv.org/abs/1801.02072}{{\tt arXiv:1801.02072}}.

\bibitem{IAXO:2013len}
{\bf IAXO} Collaboration, I.~Irastorza et~al., {\it {The International Axion
  Observatory IAXO. Letter of Intent to the CERN SPS committee}}, .

\bibitem{Armengaud:2014gea}
E.~Armengaud et~al., {\it {Conceptual Design of the International Axion
  Observatory (IAXO)}},  {\em JINST} {\bf 9} (2014) T05002,
  [\href{http://arxiv.org/abs/1401.3233}{{\tt arXiv:1401.3233}}].

\bibitem{IAXO:2019mpb}
{\bf IAXO} Collaboration, E.~Armengaud et~al., {\it {Physics potential of the
  International Axion Observatory (IAXO)}},  {\em JCAP} {\bf 06} (2019) 047,
  [\href{http://arxiv.org/abs/1904.09155}{{\tt arXiv:1904.09155}}].

\bibitem{IAXOwhite:2020}
J.~vogel~et al., {\it {Letter of Interest Snowmass 2021: The International
  Axion Observatory (IAXO) and BabyIAXO: Next Generation Helioscope Search for
  Axion and ALP Dark Matter}},  {\em .}

\bibitem{Abeln:2020ywv}
{\bf BabyIAXO} Collaboration, A.~Abeln et~al., {\it {Conceptual Design of
  BabyIAXO, the intermediate stage towards the International Axion
  Observatory}},  \href{http://arxiv.org/abs/2010.12076}{{\tt
  arXiv:2010.12076}}.

\bibitem{Ge:2020zww}
S.-F. Ge, K.~Hamaguchi, K.~Ichimura, K.~Ishidoshiro, Y.~Kanazawa, Y.~Kishimoto,
  N.~Nagata, and J.~Zheng, {\it {Supernova-scope for the Direct Search of
  Supernova Axions}},  {\em JCAP} {\bf 11} (2020) 059,
  [\href{http://arxiv.org/abs/2008.03924}{{\tt arXiv:2008.03924}}].

\bibitem{Baryakhtar:2018doz}
M.~Baryakhtar, J.~Huang, and R.~Lasenby, {\it {Axion and hidden photon dark
  matter detection with multilayer optical haloscopes}},  {\em Phys. Rev. D}
  {\bf 98} (2018), no.~3 035006, [\href{http://arxiv.org/abs/1803.11455}{{\tt
  arXiv:1803.11455}}].

\bibitem{BREAD:2021tpx}
{\bf BREAD} Collaboration, J.~Liu et~al., {\it {Broadband Solenoidal Haloscope
  for Terahertz Axion Detection}},  {\em Phys. Rev. Lett.} {\bf 128} (2022),
  no.~13 131801, [\href{http://arxiv.org/abs/2111.12103}{{\tt
  arXiv:2111.12103}}].

\bibitem{Fan:2024mhm}
X.~Fan, G.~Gabrielse, P.~W. Graham, H.~Ramani, S.~S.~Y. Wong, and Y.~Xiao, {\it
  {Highly Excited Electron Cyclotron for QCD Axion and Dark-Photon Detection}},
   \href{http://arxiv.org/abs/2410.05549}{{\tt arXiv:2410.05549}}.

\bibitem{Aja:2022csb}
B.~Aja et~al., {\it {The Canfranc Axion Detection Experiment (CADEx): search
  for axions at 90 GHz with Kinetic Inductance Detectors}},  {\em JCAP} {\bf
  11} (2022) 044, [\href{http://arxiv.org/abs/2206.02980}{{\tt
  arXiv:2206.02980}}].

\bibitem{Schutte-Engel:2021bqm}
J.~Sch\"utte-Engel, D.~J.~E. Marsh, A.~J. Millar, A.~Sekine, F.~Chadha-Day,
  S.~Hoof, M.~N. Ali, K.-C. Fong, E.~Hardy, and L.~\v{S}mejkal, {\it {Axion
  quasiparticles for axion dark matter detection}},  {\em JCAP} {\bf 08} (2021)
  066, [\href{http://arxiv.org/abs/2102.05366}{{\tt arXiv:2102.05366}}].

\bibitem{Buschmann_2022}
M.~Buschmann, C.~Dessert, J.~W. Foster, A.~J. Long, and B.~R. Safdi, {\it Upper
  limit on the qcd axion mass from isolated neutron star cooling},  {\em
  Physical Review Letters} {\bf 128} (Mar., 2022).

\bibitem{Cowan:2010js}
G.~Cowan, K.~Cranmer, E.~Gross, and O.~Vitells, {\it {Asymptotic formulae for
  likelihood-based tests of new physics}},  {\em Eur. Phys. J. C} {\bf 71}
  (2011) 1554, [\href{http://arxiv.org/abs/1007.1727}{{\tt arXiv:1007.1727}}].
  [Erratum: Eur.Phys.J.C 73, 2501 (2013)].

\end{thebibliography}\endgroup
\end{document}